\documentclass[sigconf, nonacm]{acmart}
\pdfoutput=1

\usepackage{hyperref}
\usepackage{cleveref}
\usepackage{etoolbox}
\newtoggle{fullversion}
\newcommand{\sys}{\textsc{ZenDB}\xspace}
\togglefalse{fullversion}
\usepackage{adjustbox}
\usepackage{caption}
\usepackage{scalerel}
\usepackage{graphicx}
\usepackage{epstopdf}
\usepackage{xcolor}
\usepackage{tcolorbox}
\usepackage{amsmath,amsfonts}
\usepackage{ifsym}
\usepackage{mdframed}
\usepackage{listings} 

\renewcommand{\baselinestretch}{0.99}

\usepackage{subfigure}
\tcbset{before={\par\pagebreak[0]\smallskip\parindent=0pt},after={\par\noindent}}
\usepackage{tree-dvips}
\usepackage{multirow}

\usepackage{qtree}
\usepackage[ruled,linesnumbered,vlined]{algorithm2e}
\usepackage{algorithmic}
\usepackage{textcomp}
\usepackage{tikz}
\newcommand*\circled[1]{\tikz[baseline=(char.base)]{
    \node[shape=circle,black,draw,inner sep=0.5pt] (char) {#1};}}
\usetikzlibrary{positioning, shapes.geometric, arrows}
\usetikzlibrary{shapes, arrows.meta, fit, shapes.geometric, fit, backgrounds}
\definecolor{codegreen}{rgb}{0,0.6,0} 
\definecolor{darkblue}{rgb}{0.0, 0.0, 0.5}
\definecolor{darkblueilp}{RGB}{0, 0, 175}
\definecolor{darkred}{rgb}{0.5, 0.0, 0.0}
\definecolor{darkgreen}{rgb}{0.0, 0.5, 0.0}
\definecolor{lightgray}{gray}{0.95}
\usepackage{tcolorbox}
\tcbuselibrary{listings, skins}
\newtcblisting{mypython}{
  left=1mm,    
  right=1mm,   
  top=0mm,     
  bottom=0mm,  
  listing only,
  listing options={
    language=Python,
    basicstyle=\ttfamily\tiny,
    keywordstyle=\color{darkblue},
    stringstyle=\color{darkred},
    commentstyle=\color{darkgreen},
    breaklines=true,
    breakindent=0pt,
    showstringspaces=false,
    lineskip=-1pt
  },
  boxrule=0.5pt,
  colback=lightgray, 
  colframe=black,    
}
\usepackage[disable]{todonotes}

\newcommand{\topic}[1]{\vspace{.5pt} \noindent{\bf #1.}}
\newcommand{\subtopic}[1]{\vspace{.5pt} \noindent{\em #1.}}

\newif\ifshowblock

\newcommand{\techreport}[1]{#1}
\newcommand{\techreportsp}[1]{#1}
\newcommand{\papertext}[1]{}
\showblocktrue

\newcommand{\squishlist}{
	\begin{list}{$\bullet$}
		{
			\setlength{\itemsep}{0pt}
			\setlength{\parsep}{1pt}
			\setlength{\topsep}{1pt}
			\setlength{\partopsep}{0pt}
			\setlength{\leftmargin}{1.5em}
			\setlength{\labelwidth}{1em}
			\setlength{\labelsep}{0.5em} } }
	
\newcommand{\squishend}{\end{list}}
\usepackage{multirow}

\usepackage{amsthm}

\newcommand\vldbdoi{XX.XX/XXX.XX}

\newcommand\vldbavailabilityurl{URL_TO_YOUR_ARTIFACTS}

\lstdefinestyle{SQLStyle}{
  language=SQL,
  showspaces=false,
  basicstyle=\ttfamily\footnotesize,
  commentstyle=\color{gray},
  mathescape=true,
  numbers=none,
  escapeinside={^}{^},
  captionpos=b,
  mathescape=false,
}

\begin{document}
\title{Towards Accurate and Efficient Document Analytics \\ with Large Language Models}

\author{Yiming Lin$^1$, Madelon Hulsebos$^1$, Ruiying Ma$^2$, Shreya Shankar$^1$, Sepanta Zeighami$^1$,  \\Aditya G. Parameswaran$^1$, Eugene Wu$^3$}
\affiliation{%
$^1$UC Berkeley, $^2$Tsinghua University, $^3$Columbia University \\
\{\url{yiminglin, madelon, shreyashankar, zeighami,  adityagp}\} \url{@ berkeley.edu}
\country{}\\
\url{mry21@mails.tsinghua.edu.cn},   \url{ewu @ cs.columbia.edu}}

\renewcommand{\shortauthors}{}
\renewcommand{\shorttitle}{}

\begin{abstract}
Unstructured data formats account for 
over 80\% of the data currently stored, 
and extracting value from such formats
remains a considerable challenge.
In particular, current approaches 
for managing unstructured 
documents do not
support ad-hoc analytical queries
on document collections.
Moreover, Large Language Models (LLMs)
directly applied to the documents themselves, or 
on portions of documents 
through a process of Retrieval-Augmented Generation (RAG),
fail to provide high-accuracy query results,
and in the LLM-only case, additionally incur high costs.
Since many unstructured documents
in a collection often follow similar templates that impart
a common semantic structure,
we introduce \sys, a document
analytics system that leverages this semantic structure,
coupled with LLMs, to answer ad-hoc SQL queries on document collections. 
\sys efficiently extracts semantic hierarchical structures from such 
templatized documents and introduces
a novel query engine that leverages 
these structures for accurate and cost-effective query execution.
Users can impose a schema on their documents, and query it,
all via SQL. 
Extensive experiments on three real-world document collections 
demonstrate \sys’s benefits, 
achieving up to 30$\times$ cost savings 
compared to LLM-based baselines, while maintaining or improving
accuracy, and 
surpassing RAG-based baselines by up to 61\% in precision and 80\% in recall, 
at a marginally higher cost. 
\end{abstract}

\maketitle




\section{Introduction}
\label{sec:introduction}

\begin{figure}[tb]
    \centering
    \includegraphics[width=1\linewidth]{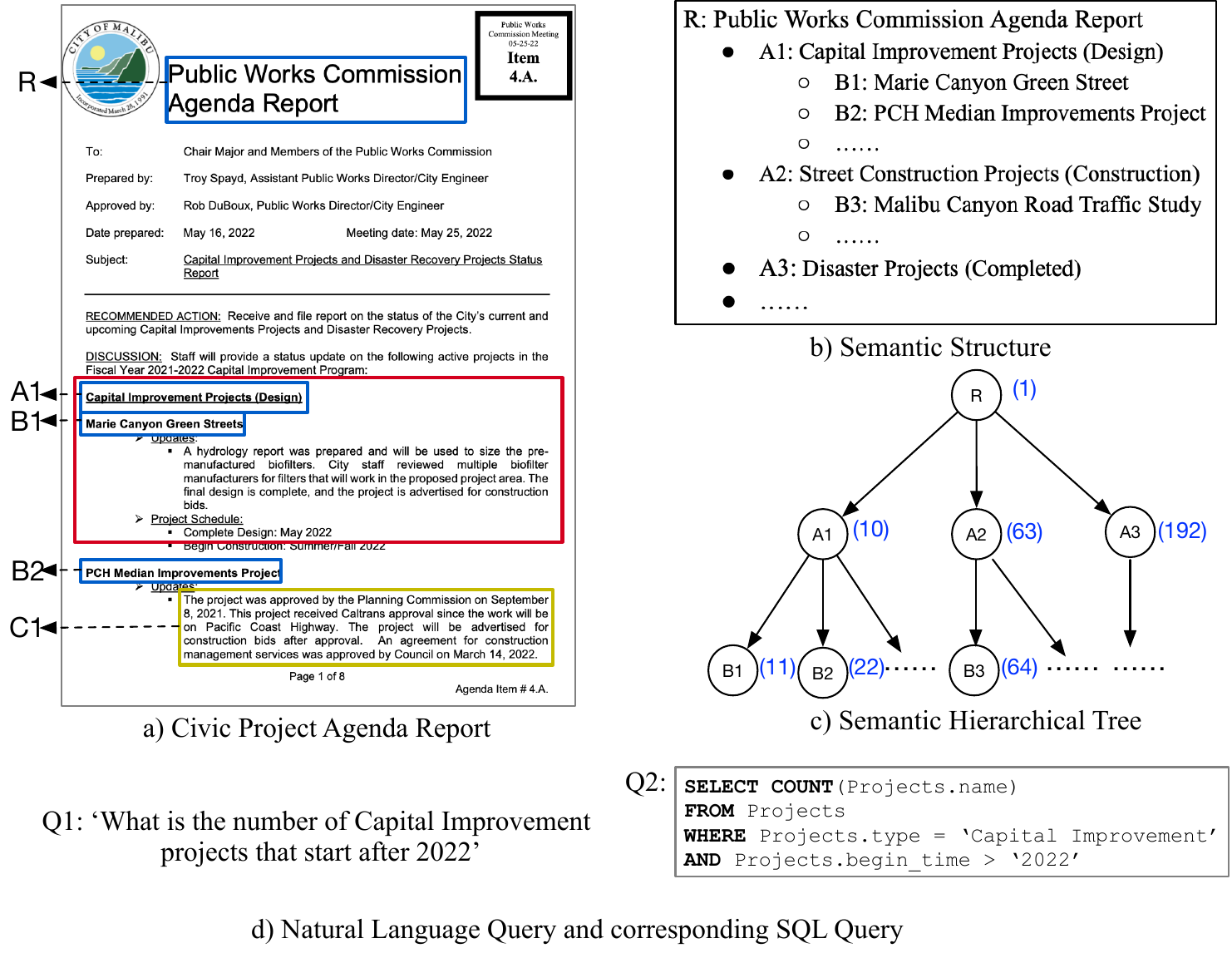}
    \vspace{-2em}
    \caption{\small Civic Agenda Document and Semantic Structures. 
    }
    \label{fig:civic}
\end{figure}




\begin{figure*}
\centering
\begin{minipage}{0.3\textwidth}
\centering
\includegraphics[width=0.9\linewidth]{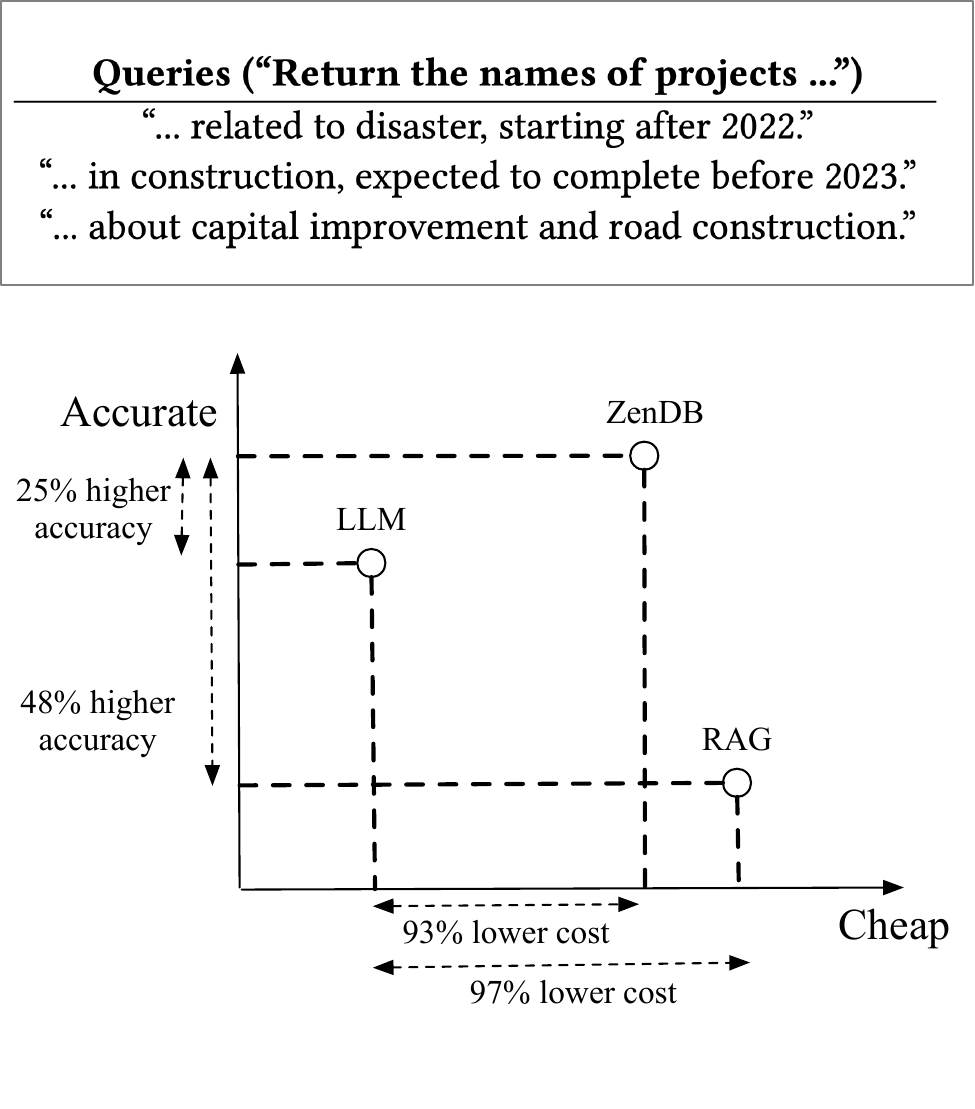}
 \vspace{-2em}
\caption{\small Understanding the differences between \sys, LLMs and RAG. }
 \vspace{-1em}
    \label{fig:accu_cost}
\end{minipage}
 \hfill
\begin{minipage}{0.66\textwidth}
\centering
\vspace{-2em}
\includegraphics[width=1\linewidth]{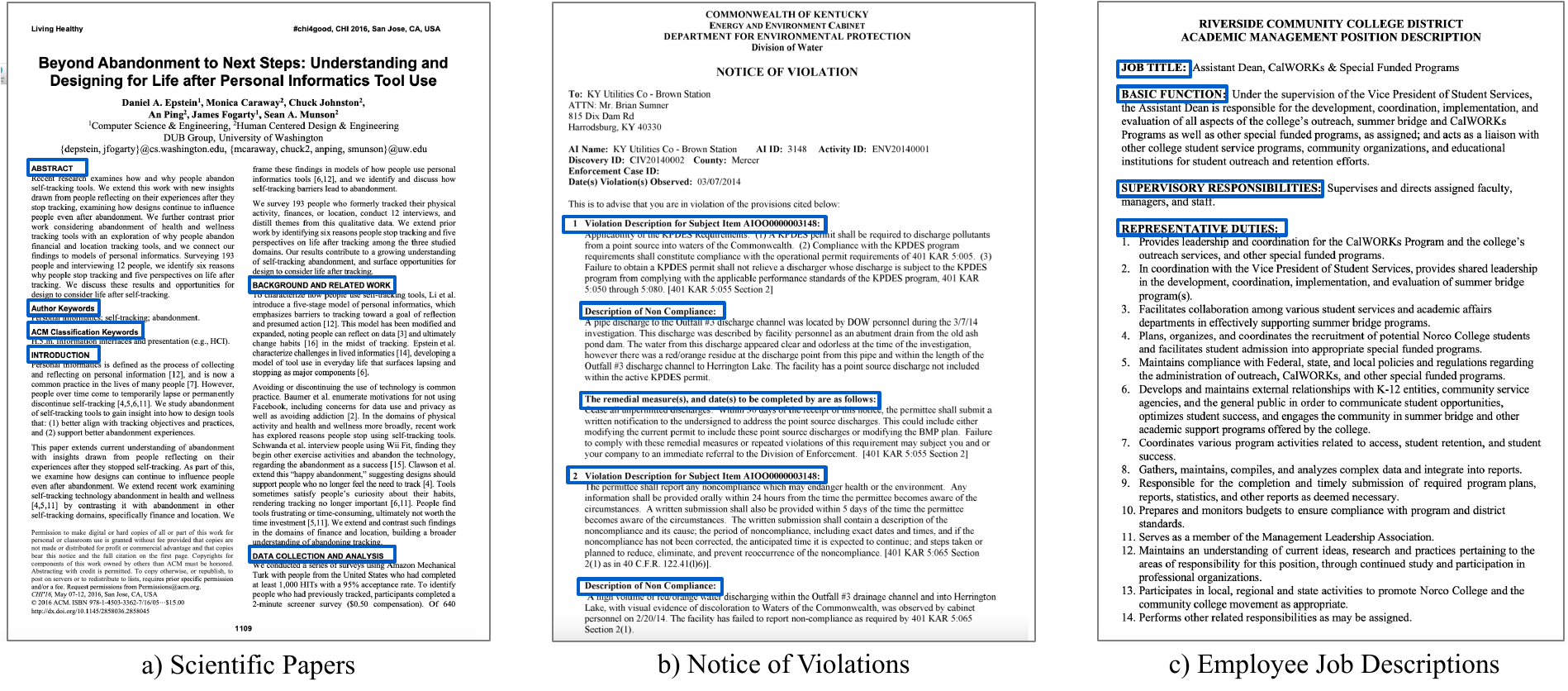}
\vspace{-1em}
\caption{\small Templatized Documents: Scientific Papers, Notice of Violations, Job Descriptions.}
    \vspace{-2.2em}
    \label{fig:data_example}
\end{minipage}
\end{figure*}


The vast majority---over 80\%---of data today 
exists in unstructured formats
such as text, PDF, video, and audio, 
and is continuing to grow at 
the rate of over 50\% annually~\cite{forbes,MITsloan}. 
In fact, an overwhelming 95\% of businesses have recognized 
management of this unstructured data as a significant problem~\cite{forbes1}. 
Consider {\em unstructured text documents},
such as Word or PDF documents, with a rich
treasure trove of untapped information.  
Due to the inherently free-form
nature of natural language, coupled with visual formatting, 
real-world unstructured documents pose a particularly  
difficult challenge for data management.
{\em Is there any hope for successfully 
querying or extracting value from unstructured documents?}

\begin{example}[Civic Agenda Report: Vanilla LLMs and RAG]
Our journalism collaborators at 
Big Local News at Stanford 
have collected
large tranches of 
civic meeting agenda PDF reports 
for various US counties as part of their agenda watch project, as in Figure~\ref{fig:civic}-a,
and want to analyze these reports.
One such query could be to count the number of
construction projects of a certain type, across meetings.
To do so, one could use Large Language Models (LLMs). 
However, even advanced LLMs, such as GPT-4, 
struggle with
queries 
issued on such reports (e.g., $Q1$ in Figure~\ref{fig:civic}-d), 
especially when these queries involve 
aggregations and/or multiple filters
on long documents. 
The error-prone nature of LLMs
is not surprising given that LLMs can't effectively handle 
large contexts~\cite{liu2024lost,bai2023longbench},
or complex data processing tasks~\cite{parameswaran2023revisiting,narayan2022can}.
The costs of processing all documents in a collection via LLMs (e.g., through OpenAI APIs)
are also high.
Another strategy, 
Retrieval-Augmented 
Generation (RAG)~\cite{lewis2020retrieval,li2022survey}, 
identifies one or more text segments within
each document that are most relevant 
(e.g., via embedding distance) 
to the given query, incorporating these segments
into prompts, reducing the cost. 
However, 
RAG struggles to identify the appropriate 
text segments, even for simple queries. 
Suppose we want to identify the
capital improvement projects. 
RAG retrieves the segments that most closely 
matches "capital improvement projects" within the document, 
such as the red box in Figure~\ref{fig:civic}-a. 
However, it fails to capture over 20 
additional projects 
in subsequent pages, such as the 
"PCH Median Improvement Project" (B2 in Figure~\ref{fig:civic}-b) 
belonging to "Capital Improvement Projects" (A1). 
Overall, both the vanilla LLM approach
and RAG are unsuitable:
both have low accuracy, while the LLM approach
additionally has high cost. 
\end{example}

\topic{Leveraging Semantic Structure Helps}
 The reason RAG didn't perform well above
was because the text segment provided
to the LLM did not leverage
the semantic structure underlying the document. 
Instead, if we are aware of this semantic structure, 
we can identify the capital improvement projects
(A1 in Figure~\ref{fig:civic}-b) 
by checking all of the subportions (e.g., B1, B2)
under it, where each one corresponds to 
the description of such a project,
and provide this
to an LLM to interpret. 
By doing so, we {\em provide all of the pertinent information
to an LLM, unlike RAG, while also not overwhelming} it with too much information.
Indeed, when we leverage semantic structure 
for a group of sample queries on GPT-4-32k, as in our system \sys, 
described next, we surpass
the vanilla LLM and RAG approaches {\bf \em by 25\%
and 48\% in accuracy, while only having 7\% of the cost} of LLMs, as detailed in Figure~\ref{fig:accu_cost}. 

\topic{Templatized Documents Provide Semantic Structure} 
Given that semantic
structure is helpful, {\em how do 
we extract this semantic structure within unstructured documents?}
Turns out, while unstructured documents 
vary considerably in format, many documents
that are part of collections are created using templates, 
which we call \textit{templatized documents}.  
Templatized documents are observed across domains,
including civic agenda reports, scientific papers, 
employee job descriptions, and notices of violations, 
as listed in Figure~\ref{fig:civic} and Figure~\ref{fig:data_example}.  
For instance, two scientific papers 
from the same venue use similar templates, 
just as civic documents for the same purpose 
from the same local county 
often adhere to a uniform template. 
Templatized documents often exhibit consistent 
visual patterns in headers (e.g., font size and type), 
when describing content corresponding to the same semantic ``level'' 
(e.g., section headers in a paper often follow the same visual pattern.)  
We highlight the ``templates'' 
using blue boxes in Figure~\ref{fig:data_example}. 
Thus, templatized documents are 
often have a discernible hierarchical
structure that reflects different semantic 
levels within the document. 
For example, a 9-page complex civic agenda report 
(such as Figure~\ref{fig:civic}-a) 
can be broken down into portions
(e.g., A1, A2, A3 in Figure~\ref{fig:civic}-b) 
and further into subportions (e.g., B2), 
indicating a possible semantic hierarchy, such as Figure~\ref{fig:civic}-c,
across the documents following the same template. 

\topic{Leveraging Semantic Structure: Challenges} 
Unfortunately, the semantic structure of the templates
isn't known---and neither do we expect 
these templates to be rigidly adhered to, 
nor do we expect there to just be one template across
the collection of documents from a specific domain.
Uncovering possible
common semantic structures across documents
is a challenge.
In addition, to support queries over unstructured
data where there isn't a predefined schema,
it's not entirely clear
what the data model or query interface should look like. 
Furthermore, using LLMs for query evaluation
incurs high monetary costs and latencies; it's
not obvious how we can leverage the semantic structures across 
documents to enable accurate query execution with low cost and latency.

\topic{Addressing Challenges in \sys}
We introduce \sys, a document analytics system
that supports ad-hoc advanced SQL queries on
templatized document collections,
and address the aforementioned challenges.
First, we introduce the notion of {\em Semantic Hierarchical Trees (SHTs)}
that represent the semantic structure for a given document,
and effectively act as an index to retrieve only portions 
of the document that are pertinent to a given query.
We build SHTs across documents 
by leveraging the uniform visual patterns 
in the document templates. 
We cluster the visual patterns found across
documents to extract and detect various template 
instantiations, coupled with minimal LLM calls
for this purpose. 
We show that if documents obey
a property we term {\em well-formattedness},
then our procedure correctly recovers their semantic structure.
Second, we introduce an extension to SQL
to query unstructured documents 
(e.g., $Q1$ in Figure~\ref{fig:civic} could be 
expressed as a SQL query $Q2$.) 
Users can easily impose a schema
on a collection of documents
by simply listing a table name as well
as a description for the entities in the table,
without listing the attributes,
which can then be lazily defined and populated
in response to queries.
Finally, we introduce a novel tree search 
algorithm that leverages SHTs to minimize cost and latency 
while answering queries
without compromising on quality. 
Specifically, we propose a summarization 
technique to create summary sketches for each node within the tree. 
\sys can navigate through the tree, 
identifying the appropriate node to answer 
a given query by examining these sketches, 
akin to how a person might use a table of 
contents to find the right chapter for a specific task.

\topic{Other Related Work}
Supporting queries on non-relational data isn't new.
For unstructured
data, the field of Information Retrieval (IR)~\cite{kobayashi2000information,singhal2001modern}
investigates the retrieval of documents 
via keyword search queries,
but doesn't consider advanced analytical queries.
For semi-structured 
data~\cite{abiteboul1997querying,mchugh1997lore,abiteboul2000data}, 
query languages
like XQuery or XPath,
as well as extensions to relational databases
for querying XML and JSON, help query hierarchically
organized data, as in our SHTs, but
there, the hierarchy is explicit
rather than implicit as in our setting.
Recent efforts have sought to bridge 
the gap between structured queries, like SQL, 
and unstructured documents.  
One line of work~\cite{urban2023towards,wu2021text} 
has explored the upfront transformation of text documents 
into tables. 
Doing this ETL process
with Large Language Models (LLMs) like GPT-4 on entire documents
is expensive and error-prone relative to approaches that 
focus the LLM's attention on specific semantic portions,
as we saw above.  
Others~\cite{thorne2021database,thorne2021natural,chen2023symphony}  have explored writing SQL queries directly on text data, 
as part of multi-modal databases. 
Most work there boils down to applying LLMs  to the entire document, 
and only works well on simple, small documents. 
However, using these methods on complex, large documents we saw above leads to high costs and reduced accuracy. 
None of the approaches above have explored 
the use of semantic structure 
to reduce cost and improve accuracy when querying documents. 
We cover this and other related work in Section~\ref{sec:related}.

We make the following contributions in this paper, as part of building \sys, our
document analytics system.
\squishlist
    \item We identify that we can leverage templates within document
    collections to support ad-hoc analytical queries.
    \item We introduce the notion of Semantic Hierarchical Trees (SHTs) that represents a
    concrete instantiation of a template for a specific document, as well as novel methods to efficiently extract 
    SHTs from an array of templatized documents.
    \item We develop a simple extension to SQL to declare a schema, 
    specify attributes on-demand, and perform analytical queries.
    \item We design a query engine that leverages SHTs, facilitating query execution in a cost-effective, efficient, and accurate manner.
    \item We implement all of these techniques within \sys and evaluate its performance on three real-world datasets, demonstrating substantial benefits over other techniques.
\squishend

\section{User Workflow with \sys}
\label{sec:architecture}

\begin{figure}[tb]
    \centering
    \includegraphics[width=1\linewidth]{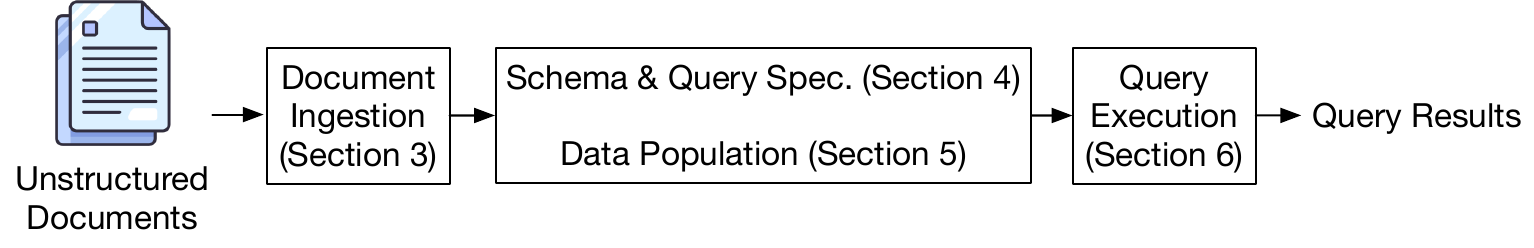}
    \vspace{-3em}
    \caption{\small User Workflow with \sys.}
    \label{fig:architecture}
    \vspace{-2em}

\end{figure}


In this section, we present an overview of 
user workflows with \sys, as illustrated in Figure~\ref{fig:architecture}. 
First, \circled{1} document collections are ingested into the system by
understanding common semantic structure (Section~\ref{sec:sht}).
Then, \circled{2} users (typically database administrators) can specify 
a schema for these documents, including
tables and lazily-specified attributes, followed
by queries that reference this schema, either specified by end-users who know SQL, or generated by applications (Section~\ref{sec:data and model}). \sys also populates upfront 
a set of system-defined tables/attributes 
to help capture the mapping between tuples and the documents (Section~\ref{subsec:datamodelpopulate}).
Finally, \circled{3} given queries on these documents, either
generated by applications or by end-users directly, \sys will execute
them efficiently, leveraging the semantic structure (Section~\ref{sec:query engine}).

\noindent\textbf{\circled{1} Semantic Structure Extraction.} 
Given a collection of templatized 
documents that adhere to one or more
predefined semantic structures, the first step within \sys
involves extracting this structure in the form of
Semantic Hierarchical Trees (SHTs), per document,
so that they can be used downstream for query execution.
This is broken down into two sub-problems:
First, how do we extract an SHT from a single document?
Second, how do we leverage common semantic structure 
across documents to scale up SHT extraction? 
Since templatized documents 
typically display consistent visual patterns 
in headers for similar semantic content, 
we cluster based on such visual patterns, 
coupled with minimal LLM invocations, 
to construct a single SHT (Section~\ref{subsec:shtsingle}). 
Then, we use a visual pattern detection approach 
to determine whether we can reuse a previously identified 
semantic structure in the form of a template, 
synthesized from a concrete SHT, 
or extract a new one (when there are multiple templates in a collection), 
all without using LLMs
(Section~\ref{subsec:shtmultiple}).

\begin{figure}[tb]
    \centering
    \includegraphics[width=0.9\linewidth]{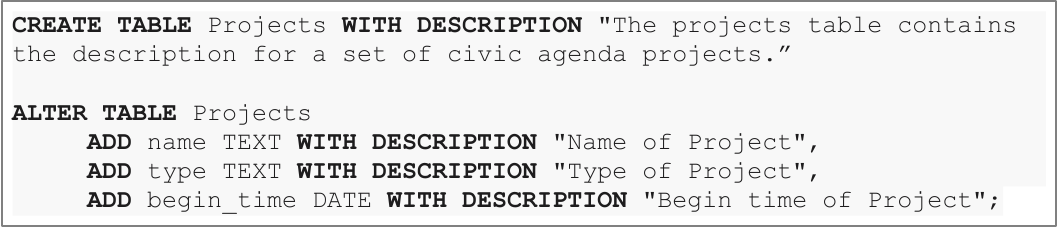}
    \vspace{-1em}
    \caption{\small Creating the Projects Table and Adding Attributes.}  
    \vspace{-2em}
    \label{fig:ddl}
\end{figure}

\noindent\textbf{\circled{2} Schema/Query Specification and Table Population. } 
Given one SHT per 
document, \sys then enables users to specify a 
schema across documents in a selection, followed by issuing queries on that schema.
Schema definition happens
via an extension of standard SQL DDL: users
(typically database administrators)
provide 
a name and description 
for each table---that we call {\em document tables}, 
along with names, types, 
and descriptions for any attributes;
the attributes can be lazily added at any point after
the table is created (Section~\ref{subsec:datamodeldef}).
For example, Figure~\ref{fig:ddl} shows
the query used to create a "Projects" table 
along with attributes (e.g., name).
Subsequently, other users
can write queries that reference such 
tables and attributes
(e.g., $Q2$ in Figure~\ref{fig:civic}),
as in standard SQL (Section~\ref{subsec:query language});
these queries could also be generated by applications
(including form-based or GUI-based applications),
or by translating natural language queries into SQL.
We still concretize the query in SQL to provide
well-defined semantics.

\vspace{1pt} \noindent 
While attributes are added lazily
and attribute values are computed or materialized
in response to queries, we proactively identify
mappings between tuples and documents during schema
specification (Section~\ref{subsec:datamodelpopulate}). 
Specifically, we identify the SHT node
that represents the portion of the document
that captures all of the relevant tuples
in a given user-specified table,
as well as the mapping between tuples
to individual SHT nodes, if they exist,
using a combination of minimal LLM invocations
and automated rules.
These are then stored in our data model as hidden
system-defined attributes, such as 
the span of the text that corresponds to the given
tuple, leveraging nodes in the SHTs built earlier.
These system-defined attributes allow for LLMs to extract
the user-defined attribute values per tuple
as needed, while reducing costs, while also leveraging
the shared semantic structure across documents.

\noindent\textbf{\circled{3} Query Execution.} 
Finally, \sys executes the user-specified
SQL queries using the pre-constructed SHTs
per document, while minimizing cost and latency, and
maximizing accuracy.
Unlike traditional relational databases, where
I/O and sometimes computation are often the bottleneck,
here,
the LLM calls invoked by \sys
becomes both a cost and latency bottleneck.
Therefore, \sys aims
to minimize such calls, while still
trying to extract attribute values as needed to answer queries,
by using a combination of predicate pushdown and projection pull-up.
We additionally develop a cost model
for \sys, focusing on monetary cost (Section~\ref{subsec:logicalplan}).
Our cost model design is flexible and can be adapted
to optimize for latency instead, e.g., if 
we instead use an open-source LLM on-prem.
Furthermore, we design novel
physical implementations
that leverage SHTs (Section~\ref{subsec:physicalplan}).
In particular, we maintain a
sketch for each node in each SHT,
and leverage this sketch
as part of a tree search
to identify
the appropriate text span 
to evaluate a given query,
akin to how a person would
use a table of contents to find the right chapter.
Finally, we maintain provenance (i.e., the specific
document text span) for query answers,
ensuring that users can verify the source
of the information and ensuring trust in the system outputs.

\begin{figure}[tb]
    \centering
    \includegraphics[width=0.95\linewidth]{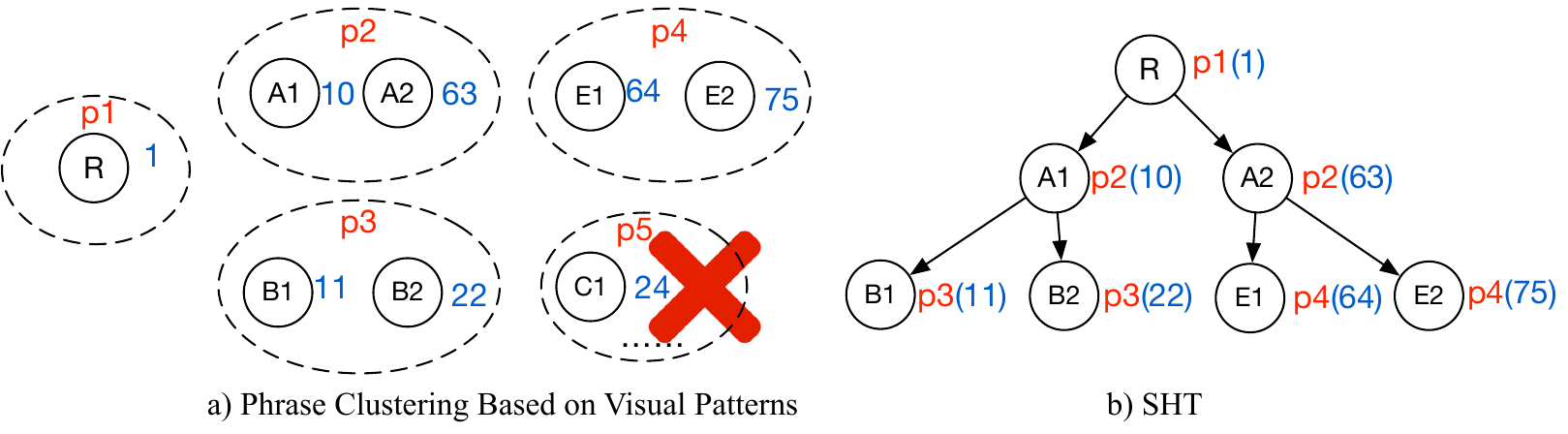}
    \vspace{-1em}
    \caption{\small SHT Construction in Civic Agenda Report.}
    \label{fig:shtbuild}
    \vspace{-0.2em}
\end{figure}

\section{Semantic Hierarchical Tree}
\label{sec:sht}

In this section, we describe our process
for recovering structure from documents in the
form of Semantic Hierarchical Trees (SHTs),
which then acts as an index for subsequent querying.
We start by formalizing the notion of SHTs and templates,
and then describe how to extract an SHT for a single document,
followed by extracting them across collections by
leveraging shared templates.

\subsection{Preliminaries}\label{sec:preliminaries}
We focus on rich text documents,
such as PDF and Word documents,
that include visual formatting information 
(e.g., multiple font types and sizes),
as shown in Figure~\ref{fig:data_example}.

\topic{Documents, Words, and Phrases}
Consider a set of documents 
$\mathcal{D} = \{D_1, D_2, ..., D_l\}$. 
For each document $D\in \mathcal{D}$,
which may be a PDF or Word document,
we often instead operate on a plain text
serialized representation,
extracted as a preprocessing step.
To generate this representation for a document $D$,
we use an extraction tool such as pdfplumber~\cite{pdfplumber},
which generates a sequence of words $W_D = [w_1, ..., w_m]$, each with 
formatting and location features (e.g., font name/size/bounding boxes).
For simplicity, we ignore images, but they
can be treated as a special word.
For any two consecutive words $w_i$ and $w_{i+1}$, if they have the same 
formatting features: font size, name (e.g., Times New Roman), 
and type (e.g., bold or underline),  
we group them into a phrase $s$. 
We let $S_D = [s_1, ..., s_n]$ be the sequence of phrases
corresponding to $D$---we often operate on $S_D$ 
instead of the document directly.

\topic{Visual Patterns}
For each phrase $s \in S_D$, we further define a {\em visual pattern},
$p(s)$, as a vector of visual formatting
features; we currently use:
$p(s) = [size, name,  type, all\_cap, num\_st, alpha\_st, center]$
but other features may be included.
Here, the first three features correspond to the font,
as in the word-level features we had previously,
and the remaining three features are phrase-level features:
$all\_cap$ is a Boolean value that denotes
whether the phrase $s$ 
is capitalized,  
$num\_st$ and $alpha\_st$ 
indicate whether the phrase starts with a number (e.g., 1) 
or a letter (e.g., A),
while 
$center$ indicates if a phrase is in the center of a line. 

\topic{Candidate SHTs} 
We are now in a position to define SHTs.
We define a {\em candidate SHT} for a document $D$
to be a single-rooted, ordered, fully connected, directed tree \(T = (V,E)\),
where each $v \in V$ corresponds to a single distinct phrase $s_i \in S_D$,
denoted $ind(v) = i$, the {\em phrase index} for $v$, 
satisfying (1) $ind(v)<ind (v')$ for any children $v'$ of $v$,
and (2) $ind(v)<ind (v')$ for any right siblings $v'$ of $v$.
These two properties together imply 
that a pre-order traversal of $T$ visits
nodes in increasing phrase index order.
A candidate SHT for
Figure~\ref{fig:civic}a is shown in Figure~\ref{fig:shtbuild}b.
Node A1 represents
the phrase (and section header)
``Capital Improvement and Disaster Recovery Projects (Design)'',
while B2 represents the phrase (and subsection header)
``PCH Median Improvement Project''.
The phrase index for each node is shown in 
parenthesis, e.g., $ind($A1$) = 10$;
i.e., A1 corresponds to $s_{10}$; ignore the p$_i$ (in red) for now.
The SHT obeys the two conditions
listed, e.g., A1 (with phrase index 10) has children (11 and 22)
and a sibling (63) with larger phrase
indexes. 

Note, however, that not all phrases in $S_D$
are found in the SHT;
this is by design: the SHT simply represents
the phrases corresponding to the
{\em headers} of the document, while those
that correspond to the {\em content} are omitted. 
For example, Figure~\ref{fig:shtbuild}b omits phrases $s_2,.. ,s_9$.
However, in certain cases, it may be convenient
to refer to headers
and content together. 
For this, we define 
{\em text span} or $ts$, to be a sequence of phrases $s_i, ..., s_{i+k} \in S_D$,
or equivalently $[i, i+k]$.
We define $next(v)$ for a given node
$v$ to be the phrase index corresponding to
its sibling to the immediate right, if available,
or, if not, the sibling to the immediate right of the closest ancestor
that has one. If none of the ancestors of $v$
have a right sibling, $next(v) = n$, where $n$ is the total
number of phrases in $S_D$.
To illustrate, $next($A1$) = next($B2$) =$ 63 (i.e., A2),
while $next($A2$)= next($R$) = 100$, assuming $s_{100}$ is 
the final phrase in our document.
A given node $v \in V$ has a text span:
$ts(v) = [ind(v), next(v)-1]$,
i.e., $v$ ``covers'' all of the phrases until the next node with
phrase index $next(v)$.
Thus, $ts($R$)$ is $[1, 100]$,
while $ts($B2$)$ is $[22, 62]$.
That is, B2 ``covers'' both the header, $s_{22}$,
as well as the content $s_{23}, \ldots, s_{62}$, until
the next header, A2.
In the following, we equivalently
refer to a node $v$, its header phrase $s_{ind(v)}$ (i.e., 
the header corresponding to $v$),
or text span $ts(v)$ (i.e., the header and content contained
within $v$).
We finally introduce the notion of a {\em granularity} or {\em height} of a node $v$,
which is simply the depth of $v$ in the SHT;
in our example, the depth of R is 1, and A1 is 2.

\subsection{SHT Construction on a Single Document}
\label{subsec:shtsingle}

Given a document $D$ with phrases $S_D$,
there are exponentially many candidate SHTs;
our goal is to identify the {\em true SHT}
that correctly reflects
the semantic structure of the document.
To do so, our procedure, 
\texttt{oracle\_gen}($D$), first identifies which phrases
are header phrases (and therefore correspond
to SHT nodes). We then
assemble these phrases into a tree,
ensuring that it is a candidate SHT.

\topic{Header Phrase Identification}
To identify if a phrase $s\in S_D$ is a header phrase,
we make use of visual patterns $p(s)$.
We cluster the phrases in $S_D$
based on their visual patterns.
For our running example,
the clusters that emerge are shown in Figure~\ref{fig:shtbuild}a,
each labeled with its visual pattern (in red).
Here, the majority of the phrases
end up in the cluster with pattern p5---this corresponds
to the content phrases in the document 
(e.g., C1 in Figure~\ref{fig:civic}-a is a paragraph).
To remove clusters whose phrases
do not correspond to header phrases,
we use LLMs as an oracle.
We randomly sample $min(|C|,k)$ ($k$ is a 
predefined threshold) 
phrases in each cluster $C \in \mathcal{C}$. 
For each sampled phrase $s\in C$, 
we construct the LLM prompt 
``Is the phrase [s] a header in the document?''. 
If over half of the sampled phrases 
in $C$ are non-headers, then $C$ is pruned 
(e.g., the cluster containing C1 is dropped
since C1 is a paragraph).  
To verify if GPT-4 is effective at
disambiguating headers from non-headers,  
we carefully examined over 200 documents 
from 16 datasets, covering six diverse domains. 
In our testing, when $k=10$, GPT-4 effectively 
removes non-header clusters on 97\% 
of the documents with total cost as \$0.37.
Still, since this cost is non-zero,
we would want to minimize it
when working on a large collection of documents;
as we illustrate in our next section,
we only invoke LLMs for a small
subset of documents,
each corresponding to a different template.

\topic{Tree Construction}
Given the header phrases across the
remaining clusters in $\mathcal{C}$, 
we assemble the corresponding
nodes into a tree. 
We proceed top-down, operating on one
cluster at a time, adding the entire cluster
to the partially constructed SHT.
At each step, we pick the cluster $C$ that
contains the phrase with the lowest index.
For each phrase $s_i$ in this cluster $C$, 
we create a corresponding node $v_i$ 
and add it to the partially constructed SHT,
in increasing phrase index order, simultaneously.
For each such node $v_i$, we examine the $ts$ of all
existing nodes in the partially constructed
SHT, and pick its parent to be the
ancestor $v_j$ such that $ind(v_i) \in ts(v_j)$,
and there is no other $v_k > v_j$ such that $ind(v_i) \in ts(v_k)$.
This condition basically ensures that $v_i$ is added
under the most specific node $v_j$ that
can accommodate it. 
Once we've identified the appropriate parents
for each node in the cluster, 
we then add all of these nodes together.
The root (usually corresponding to $s_1$) 
merits special treatment:
if there is no cluster that contains $s_1$,
we create a node corresponding to $s_1$,
else we start with the cluster that contains $s_1$.
Usually this cluster just has $s_1$;
if it contains other phrases,
we create an artificial root node corresponding to 
an empty phrase $s_0$, and deem it to be the root.
We then process the cluster that contains $s_1$ along
with other phrases.
Returning to our example, the cluster corresponding
to visual pattern $p_1$ with phrase $s_1$ is processed first,
allowing R to be added to the tree.
Then, the cluster corresponding to $p_2$ is processed
next as it has the lowest phrase index number 10,
with A1 and A2 added to the tree together,
both with R as parent.
Then, the cluster corresponding to $p_3$ is processed, 
with B1 and B2 being added as children of A1,
and so on.

\topic{Correctness for Well-Formatted SHTs}
Next, we show that if the true SHT for a document
has a property that we call {\em well-formattedness},
then \texttt{oracle\_gen}($D$) correctly outputs the 
true SHT.

Given an SHT $T$, the visual prefix $vispre(v)$
for a node $v$ is defined to be the sequence
of visual patterns from the root to $v$.
In our example, $vispre($B1$) = p_1p_2$.
We extend the definition to a set in the natural
way, e.g., $vispre($\{B2, A1\}$)$ = \{$p_1, p_1p_2$\}.
Let $pset(p)$ be a
function that accepts a visual pattern and returns
all the nodes that obey that pattern. For example,
$pset(p_2) =$\{A1, A2\}.

Then, an SHT $T=(V,E)$ is said to be {\em well-formatted}
if (1) for any two siblings $v_i, v_j$, $p(v_i)=p(v_j)$;
(2) for all visual patterns $p$, $vispre(pset(p))$
is unique.
The first condition mandates that sibling nodes, 
such as $B1$ and $B2$, must share the same visual pattern. 
However, it does not require that 
all nodes at the same depth, 
like $B2$ and $E1$, must have identical visual patterns.
In our agenda watch dataset, subsection headers
within a section often have similar formatting,
but this need not hold across sections, i.e.,
different sections may use different formatting.
The second condition states that nodes sharing 
the same visual pattern must have identical visual prefixes. 
For example, $B1$ and $B2$ have the visual prefix $p_1p_2$.
Thus, the visual pattern signifies
a certain ``semantic level'' within the SHT,
following a consistent path to the root. 
\begin{theorem}
\label{theo:sht}
\vspace{-2pt}
    If the true SHT for a document $D$ is {\em well-formatted}, and if
    an LLM can correctly identify non-headers, then 
    \texttt{oracle\_gen}($D$) outputs the 
    true SHT.
\vspace{-2pt}
\end{theorem}

\papertext{\noindent We show the proof in our 
technical report~\cite{textdblong}.}

\techreportsp{
\begin{proof}}
\techreport{
Let $T$ and $GT$ be the SHT returned by \texttt{oracle\_gen} and in the ground truth, respectively. 
We prove $T=GT$ when $T$ is a well-formatted SHT by induction. Let $v_i$ be the i-th node added in the \texttt{oracle\_gen} approach, and $N_{i-1} = \{v_1,v_2,...v_{i-1}\}$ be the first $(i-1)$-th nodes added in \texttt{oracle\_gen}, respectively. Let $T_{i-1}$ and $GT_{i-1}$ be the induced subgraph of $T$ and $GT$ based on the set of nodes $N_{i-1}$. By induction, we assume that \texttt{oracle\_gen} returns the correct SHT, i.e., $T_{i-1} = GT_{i-1}$, when adding the first $(i-1)$-th nodes, and we further prove that, by adding $v_i$, $T_i = GT_{i}$. 
}

\techreport{Let $v_j$ and $v_{j}^{'}$ be the parent node of $v_i$ in $T_i$ and $GT_{i}$, respectively. We prove that $v_j = v_j^{'}$ by considering two cases: one where there exists a node $v_k \in T_{i-1}$ (and $v_k \in GT_{i-1}$, since $T_{i-1} = GT_{i-1}$) shares the same visual pattern as $v_i$, i.e., $p(v_k) = p(v_i)$, and one where it does not. Let $g(v)$ be the granularity (i.e., height of node) of $v$ in $T_i$. Let $path(v_i)$ be the sequence of nodes from root to $v_i$ in $T_i$.
}

\techreport{
Assume $\exists v_k \in T_{i-1}$ and $v_k\in GT_{i-1}$, s.t., $p(v_k) = p(v_i)$.  $g(v_j) = g(v_i)+1$ since $v_j$ is the parent node of $v_i$ in $T_i$.  
By definition, $\forall v\in path(v_i), v\neq v_i$, we have $ind(v) < ind(v_i)$ and $ind(v_i) \in ts(v)$. We call each $v\in path(v_i), v\neq v_i$ as a candidate parent node of $v_i$ since adding an edge from $v$ to $v_i$ will make $T_i$ a valid candidate SHT. Thus $v_j^{'} \in path(v_i)$ since GT should be at least a valid SHT and there is no other $v_m > v_j$ such that $ind(v_i) \in ts(v_m)$. If $g(v_j^{'}) \neq g(v_j)$, there at least exists one node $v_l\in path(v_i)$ and $v_l$ is a child node of $v_j^{'}$, s.t., $p(v_l) = p(v_i)$, since $GT_i$ is a well-formatted SHT and the sibling nodes $v_l$ and $v_i$ belonging to the same parent $v_j^{'}$ should have the same visual pattern. By $p(v_k) = p(v_i)$, we have $p(v_l) = p(v_k)$, and thus $g(v_l) = g(v_k)$, since $vispre(v_l) = vispre(v_k)$. $g(v_l) = g(v_k)$ implies $g(v_j^{'}) =g(v_j)$, which contradicts with $g(v_j^{'}) \neq g(v_j)$. By contradiction, we have $g(v_j^{'}) =g(v_j)$ and further $v_j = v_j^{j}$ since both $v_j$ and $v_j^{'}$ are in $path(v_i)$. 
}

\techreport{
Assume $\nexists v_k \in T_{i-1}$ and $v_k\in GT_{i-1}$, s.t., $p(v_k) = p(v_i)$. Similarly we show $v_j = v_j^{'}$ by contradiction in this case. Assuming $v_j \neq v_j^{'}$, there at least exist a node $v_l \in path(v_i), v_l\neq v_i$ and $v_l$ is a child of $v_j^{'}$, s.t., $p(v_l) = p(v_i)$, since $v_j^{'} \in path(v_i)$ and $GT_i$ is a well-formatted SHT. However, this contradicts to the assumption that $p(v_k) = p(v_i)$. By contradiction, we have $v_j = v_j^{'}$, which concludes the proof. 
}
\techreportsp{
\end{proof}
}


\subsection{SHT Construction across Documents}
\label{subsec:shtmultiple}

Given a set of documents 
$\mathcal{D} = \{D_1, ..., D_l\}$,
applying \texttt{oracle\_gen}($D_i$) to  each $D_i$
can be costly when $l$ is large.
Here, we leverage the fact that,
in addition to being {\em well-formatted}, 
the documents share common {\em templates}.
We define the notion of a template below.
We process each document $D_i$ in turn, 
attempting it to match against one of the existing
templates $tp \in \mathcal{TP}$
via a function \texttt{template\_gen}($tp, D_i$);
if a match is successful, a SHT for $D_i$
is returned--without any LLM calls.
Otherwise, we call \texttt{oracle\_gen}($D_i$)---here,
the corresponding template $tp$ for the returned
SHT is added to $\mathcal{TP}$.
If there are multiple successful matches in $\mathcal{TP}$,
we return the largest SHT of them all;
the rationale here is that
we want to capture as much of the header information 
as possible as part of the SHT.

\begin{figure}[tb]
    \centering
    \includegraphics[width=0.8\linewidth]{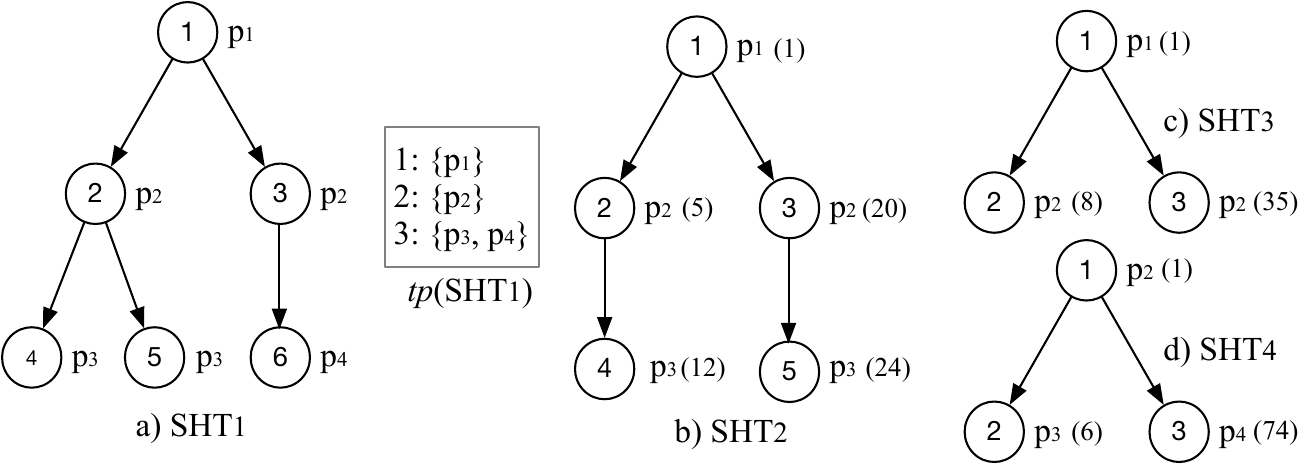}
    \caption{\small SHT construction by Pattern Matching; the documents
    represented by b and c are matches to $tp($SHT1)$)$ but not d.}
    \label{fig:sht}
\end{figure}

\begin{figure*}[tb]
    \centering
    \includegraphics[width=0.97\linewidth]{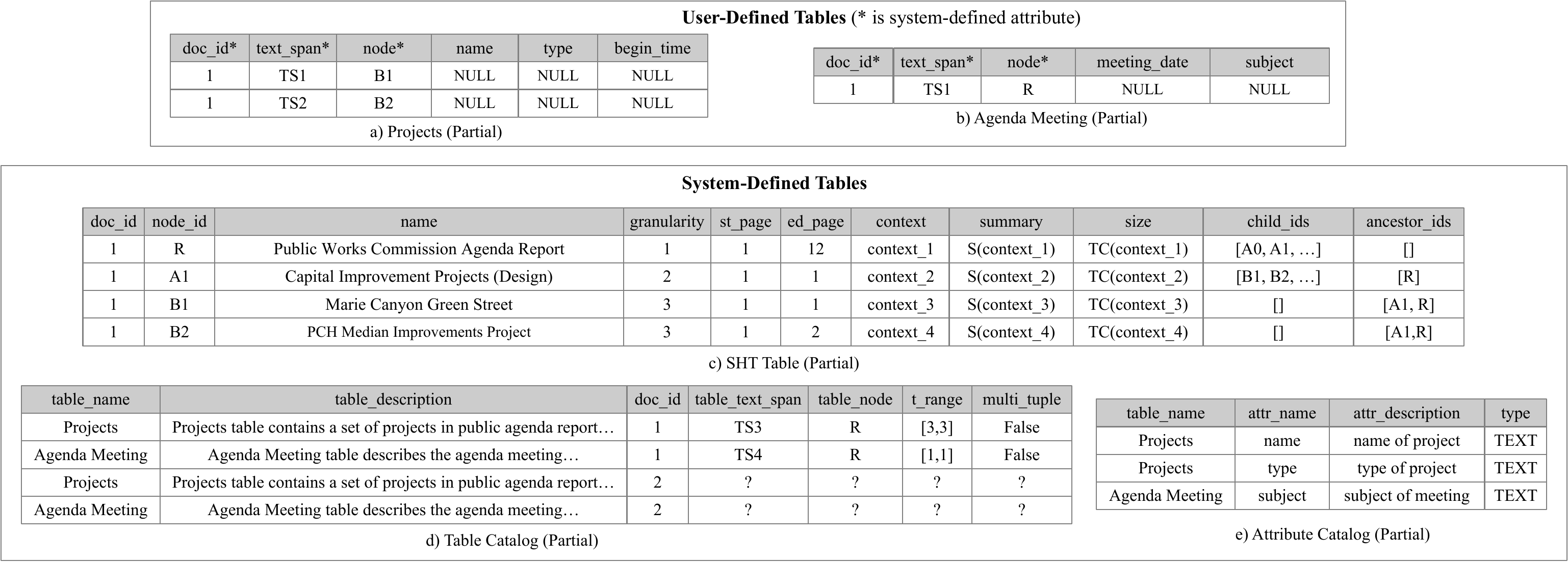}
    \vspace{-1em}
    \caption{Data Model: User-Defined Tables and System-Defined Tables. }
    \vspace{-1em}
    \label{fig:model_query}
\end{figure*}

\topic{Template}
We now define the notion of a template
associated with an SHT.
The {\em template} for an SHT $T$: 
$tp = \{g: \{p\}\}$ is 
a sorted dictionary that 
captures the mapping between 
the granularities $g$ of nodes and the set $\{p\}$
of visual patterns found at that
granularity.
This dictionary is additionally sorted by granularity in 
increasing order.
$tp($SHT1$)$ is the template of 
SHT1 shown in Figure~\ref{fig:sht}-a. 
We let $tp.g$ and $tp.p$ 
be the granularities 
and visual patterns in $tp$. 
For SHT1 in Figure~\ref{fig:sht}-a, 
$tp.g = \{1,2,3\}$ and $tp.p = \{p_1,p_2,p_3,p_4\}$. 
Let $tp.g(p)$ 
be the granularity of a visual pattern $p$ in $tp$, 
e.g., $tp.g(p_1)$ = 1 for SHT1. (This value
is unique by construction from Section~\ref{subsec:shtsingle}.)
If $p\notin tp.p$, $tp.g(p) = -1$.

\topic{Template Matching and Generation}
We say a document $D$ {\em matches} 
a template $tp$ if the visual patterns
contained amongst the phrases $S_D$
cover each granularity $1 \ldots i$,
for some $i$ which is a prefix of $tp$.
For instance, document $D_1$ with 
true SHT, SHT1, 
has a corresponding template  
$tp($SHT1$)$ in Figure~\ref{fig:sht}-a, 
and document $D_2$,
has a true SHT, SHT2, Figure~\ref{fig:sht}-b. 
Since $D_2$ includes patterns $\{p_1, p_2, p_3\}$,
it covers every granularity of the template
of $D_1$,
and therefore {\em matches} the template.
Additionally, document $D_3$ with true SHT, SHT3, in Figure~\ref{fig:sht}-c,
which includes patterns $\{p_1, p_2\}$,
also matches the template,
since it covers a prefix of the granularities
in the template (namely 1 and 2),
even though it lacks patterns $\{p3,p4\}$.
On the other hand, document $D_4$ with true SHT, SHT4, in Figure~\ref{fig:sht}-d,
does not contain a match for $p_1$,
thereby not meeting the prefix constraint, and not being 
a match for the template.
Our rationale for admitting prefix matches
is the observation that as 
the granularity of a header 
becomes more fine-grained, 
its visual pattern tends to be more varied. 
For example, for two scientific papers obeying the same template, 
the visual patterns of sections remain consistent,
but within each section the visual patterns used
may vary depending on individual preferences. 
Note here that 
in our implementation, we allow for any non-zero prefix
for a match;
for more constrained document collections,
a user may set a prefix threshold, e.g., at least three levels
of the template must be covered.

Armed with templates and matches
to a template, we can now describe our
\texttt{template\_gen}$(tp,D)$ procedure\techreport{, listed 
in Algorithm~\ref{alg:treegen}}.
We proceed in two phases,
where we first identify
all of the phrases $s \in S_D$
that match those in $tp.p$,
we add these phrases as nodes to $V$
for our yet-to-be-constructed SHT \techreport{(Line 3-5)}.
Given these phrases, we check if
there is a match for the template $tp$,
where a match is defined as above to be a prefix
of the template.
If no match is found, an empty result
is returned \techreport{(Line 6-7)}, else
we assemble the nodes in $V$ into an SHT;
we use a similar tree construction procedure
as in the previous section, operating on the 
phrases found in the first step,
clustered based on visual pattern\techreport{~(Line 8-10)}. 
\techreport{
\setlength{\textfloatsep}{0pt}
\begin{algorithm}[bt]
    \small
	\caption{\texttt{template\_gen}$(tp,D)$}
	\label{alg:treegen}
    $SHT_D = (V,E), V=\emptyset, E=\emptyset$ \\
    $G = \{\}$ \\ 
    \For{$s_i\in S_D$}{
    \If{$p(s_i) \in tp.p$}{
        $V = V\cup s_i$, $G = G \cup tp.g(p(s_i))$ \\
    }
    }
\If{$G = \emptyset$ or $\exists i \in G, i > 1, (i-1)\notin G$ }{\textbf{Return} {\{\}}\\}
\For{$v_i\in V, v_j \in V$}{
\If{$ind(v_j) \in ts(v_i)$ and $\nexists v_k\in V, ind(v_k) > ind(v_i), s.t., ind(v_j)\in ts(v_k)$}{$E = E\cup (v_i, v_j)$ \\}
}
 \textbf{Return} $SHT_D$
\end{algorithm}
}


\section{Data Model and Query Language}
\label{sec:data and model}

In the previous section, we described
how we can extract SHTs for each
document in a collection as part of document ingestion.
Here, we define the data model 
used by \sys to represent 
the SHTs as well as other system-specific information,
along with user-defined tables that we call {\em DTables},
short for {\em Document Tables}.

\subsection{Data Model Definition}
\label{subsec:datamodeldef}

In addition to 
traditional relational tables that we call {\em base tables}, 
\sys supports three new types of tables
that respectively (i) represent the SHTs per document collection,
(ii) let users 
specify one or more 
structured relations over the documents,
called {\em DTables},
to be used within queries;
(iii) maintain system metadata associated
with the user-defined tables.
We describe each one in turn.

\subsubsection{SHT Table}
The \textit{SHT table}, 
shown in \Cref{fig:model_query}-c, is 
a system-defined and maintained table 
that represents the SHTs in a document collection.
Each row captures information about an SHT Node,
and is populated as described subsequently in \Cref{subsec:datamodelpopulate}.  
Its main attributes are:
\squishlist
    \item \texttt{doc\_id}, \texttt{node\_id} identify the node in a given document.
    \item \texttt{name} represents the header phrase $s$ corresponding to the node.
    \item \texttt{granularity} represents the depth of the node in the tree.
    \item \texttt{context}, \texttt{summary}, \texttt{size} correspond to the 
    entire sequence of phrases in the text span, a short summary of the
    text span, 
    and the number of tokens in the text span.
    \item \texttt{st\_page} and \texttt{ed\_page},
    listing the start and end pages for the text span.
    \item \texttt{child\_ids} and \texttt{ancestor\_ids}, the IDs for the children
    and entire sequence of ancestors.  
\squishend
We note that \texttt{summary}, \texttt{size}, \texttt{st/ed\_page}, and 
\texttt{ancestor\_ids} can be derived from the other attributes, 
but we store them
explicitly for convenience. These attributes
are all used during query processing.

\subsubsection{User-defined DTables}
Users can use SQL to define DTables, 
with those tables 
being used 
in subsequent queries (\Cref{fig:ddl}).  
We use a special keyword \texttt{DESCRIPTION} to
both designate the fact that this 
is not an ordinary table, and also
allowing natural language to be provided 
that may be used in LLM prompts. 
To define such a table, the user can say:
\begin{lstlisting}[style=SQLStyle,mathescape,]
CREATE TABLE [name] (...) WITH DESCRIPTION [description]
\end{lstlisting}
Here, the user provides a natural language description
for the table. 
Attributes may be provided during table creation 
in parentheses (or omitted),
and/or could be added afterwards, via the
standard approach to alter schemas:
\begin{lstlisting}[style=SQLStyle,mathescape,]
ALTER TABLE [name] 
ADD [name] [type] WITH DESCRIPTION [description], ... ;
\end{lstlisting}
Again, a natural language 
description for the attributes are provided 
when they are added.
As we will discuss in \Cref{subsec:datamodelpopulate},
when the user creates a DTable, \sys populates them offline
with rows that correspond to tuples.
Each tuple represents one entity 
that can be found in a document. 
User defined attributes for these tuples are populated
with \texttt{NULL}, and 
are filled in on-demand during query time,
as shown in \Cref{fig:model_query}a.
Here, the \texttt{Project} DTable contains user-defined attributes \texttt{name}, 
\texttt{type}, and \texttt{begin-time}. 
\sys also maintains three hidden system-defined attributes
per DTable---the 
document id, text span used to extract the tuple, 
and SHT nodes used in the derivation.  
These attributes track how each tuple was derived, 
to provide context when extracting tuple attributes later on, 
and for debugging and provenance purposes. 
For instance, $B1$ corresponds to 
the ``Marie Canyon Green Street'' project tuple, 
and the tuple's text span may be the same as $B1$ 
or a subset (\Cref{fig:model_query}c). 

\techreport{
The user-defined attributes represent the result of a {\it read} 
operation over each attribute.    
In addition, every expression 
implicitly defines additional attributes in this table.  
For instance, if a query evaluates 
\texttt{Projects.name = ``Capital Improvement''} 
directly using an LLM call, then the attribute 
\texttt{[Projects.name|eq|Capital Improvement]} 
is instantiated and populated with the LLM response.}

Note that we chose to represent
these user-specified DTables as regular tables
as opposed to views or materialized views;
but they could also be represented
as such.

\subsubsection{System-Defined Tables}
In addition to
the SHT table, \sys maintains 
two system-defined tables: 
\texttt{Table Catalog} 
and \texttt{Attribute Catalog} 
store metadata related to tables and attributes 
respectively (Figure~\ref{fig:model_query}d,e).
In addition to names and descriptions, 
\texttt{Table Catalog} tracks the text span 
and SHT node(s) used to 
identify the contents of the table 
(since a table may be a small portion of the document),
used to localize search when extracting tuples---thereby reducing
cost during query processing. 
The attribute \texttt{t\_range} 
refers to the min/max 
granularities of the nodes 
used to extract tuples in the table.  
For example, all \texttt{Project}
tuples extracted so far have granularity 3, 
thus \texttt{t\_range = [3,3]};
this is the setting where tuples
correspond to nodes (of some granularity) within the SHT.  
Finally, to handle the special case
where the table is extracted
from a leaf node in the SHT, i.e., there
are multiple tuples corresponding to a single node that has no finer granularity node
below it,
we mark this 
by setting \texttt{multi\_tuple} to \texttt{True}.
For instance, 
consider the scenario when 
users want to create a table called ``References''  
and each tuple corresponds to a reference 
in a published paper.  



\subsection{Query Language}
\label{subsec:query language}

\sys currently supports
a subset of SQL,
corresponding to simple non-nested queries 
on one or more DTables 
with optional aggregation, 
as represented by the following 
template:
\begin{lstlisting}[style=SQLStyle,mathescape,]
SELECT [attr] | agg(attr) FROM [ST]+
 WHERE [predicate]    GROUP BY [attr]
\end{lstlisting}
where \texttt{[..]} denotes a list of elements, 
\texttt{attr} refers to an expression over an attribute, 
\texttt{ST} refers to one or more DTables, 
and \texttt{agg()} 
includes \texttt{SUM, COUNT, AVG, MAX, MIN}\techreport{\footnote{Text attributes only support \texttt{COUNT}, date attributes only support \texttt{COUNT, MAX, MIN}.}}.    
A predicate has the form: \texttt{attr op operand}, 
where the operators include \texttt{>|$\geq$|<|$\leq$|=|LIKE|IN}, 
and  \texttt{operand} is one or more constants.  
\texttt{LIKE} is used for fuzzy matching 
where either string similarity 
or semantic similarity 
could be used\footnote{In \sys we use 
Jaccard similarity with a 0.9 threshold by default.}.
We add a restriction that if multiple DTables
are listed in the \texttt{FROM} clause,
then the \texttt{WHERE} clause includes a predicate 
specifying that the tuples are equi-joined on \texttt{doc\_id}.
We add this restriction for now to only allow for within-document
joins, but we plan to relax this in future work.  

\begin{figure}[tb]
    \centering
    \includegraphics[width=0.8\linewidth]{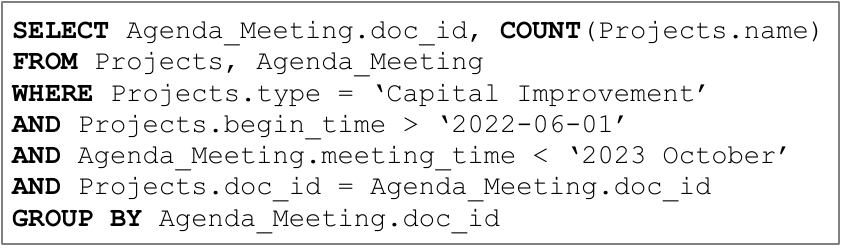}
    \vspace{-1em}
    \caption{A Query on Civic Agenda Documents.}
    \label{fig:query}
\end{figure}

%
Figure~\ref{fig:query} shows a query where, 
for each document 
whose meeting time is before ``2023 October'', 
we count the ``Capital Improvement''  
projects starting after ``2022-06-01'';
here, we make use of the within-document join
across two tables.

The query semantics are defined 
as fully populating the user-defined DTables 
with the LLM results\techreport{~of all attribute 
reads and expressions}, 
and then executing the SQL query as normal.   
We follow these semantics because 
it allows for minor consistencies during query evaluation.  
\techreport{
Specifically, under an oracle LLM that 
always returns complete and correct responses, 
the contents of the attribute reads and expressions will always be consistent 
(e.g., \texttt{type} is \texttt{['A', 'B']}, and \texttt{type = 'A'} is true).}
 However, modern LLMs are imperfect and 
 sensitive to the input prompt and context formulation, 
 so the extracted attribute values and expressions 
 over the attributes may be inconsistent 
 (e.g., extracted \texttt{type} is \texttt{'B'}, but \texttt{type='A'} is true).    
 Better understanding and reconciling these 
 potential inconsistencies is outside the scope of this paper, 
 and is important future work.

\section{Table Population}
\label{subsec:datamodelpopulate}
We next describe how 
we can populate the system-defined tables and attributes
described above. 
Populating the SHT table is straightforward
and therefore omitted;
we will describe how the \texttt{summary}
field is populated in Section~\ref{sec:query engine}.

\topic{Populating Tables Overview}
When a user defines a new DTable \texttt{T}, 
updating  
\texttt{Attribute~Catalog} (Figure~\ref{fig:model_query}e) 
and \texttt{table\_name}, \\\texttt{table\_descr} 
in \texttt{Table~Catalog} (Figure~\ref{fig:model_query}d)
is easy.  
However, \sys must process 
the document collection $\mathcal{D}$ to fill in 
the system-defined attributes (SDAs) 
in \texttt{Table~Catalog} 
and \texttt{T}, and 
populate \texttt{T} with tuples.  
While \sys proactively 
identifies tuples for \texttt{T}, 
it doesn't
populate any user-defined attributes until query time. 

Consider a partitioning of $\mathcal{D} = \bigcup_{\mathcal{D}_i \subseteq \mathcal{D}} \mathcal{D}_i$, 
where $\mathcal{D}_i$ is a set of documents 
sharing the same template, 
as identified during SHT construction. 
For each $\mathcal{D}_i$, \sys picks a document $D\in \mathcal{D}_i$ 
and uses an LLM to populate \texttt{T} with tuples, 
and fill in the SDAs.  
\sys then uses a rule-based approach to extract tuples 
from the remaining documents $D'\in\mathcal{D}_i - \{D\}$ without invoking LLMs.
We describe the single document and multi-document 
extraction next.




\topic{Single Document Extraction}
To populate SDAs for $D$ for a given DTable \texttt{T}, 
we first identify the node in the SHT
for $D$ that captures all of the entities for
the \texttt{T}; 
we call this the {\em table node}.
We then identify nodes that correspond to tuples
that lie underneath this node. 
We use two prompts, \texttt{table\_oracle} 
and \texttt{tuple\_oracle}
to identify if a given node corresponds
to a table or tuple respectively. 

\vspace{-0.3em}
\begin{mypython}
table_oracle: If the following text describes [table_name], [table_descr], return true. Otherwise, return false. [node_context]. 

tuple_oracle: If the following text describes one [tuple_descr] in [table_name], [table_descr], return true. Otherwise, return false. [node_context]. 
\end{mypython}

\noindent In these prompts, \texttt{[]} is a placeholder. 
\texttt{[table\_name]}, 
\texttt{[table\_descr]}, 
and \texttt{[tuple\_descr]} 
correspond to the table name and 
description, and the tuple description in 
\texttt{Table Catalog} (e.g., Figure~\ref{fig:model_query}d).  
\\ \texttt{[node\_context]} provides
the entire text span corresponding to the node from
\texttt{SHT table} (e.g., in Figure~\ref{fig:model_query}c).  

To identify the table node, 
\sys walks the SHT top-down and submits \texttt{table\_oracle} 
to LLMs for each node. 
If the response for all of a node $v$'s children are true, 
then we add $v$ as a candidate table node
and stop descending into $v$'s children. 
Finally, \sys fills in  
the Least Common Ancestor (LCA) of the candidate table nodes 
as \texttt{table\_node} in \texttt{Table Catalog}. 

Once the \texttt{table\_node} is found, 
\sys attempts to populate \texttt{T} with tuples. 
Once again, \sys performs a top-down traversal 
starting from \texttt{table\_node}
and evaluates \texttt{tuple\_oracle} on each node.  
If a node $v$ evaluates to true, it means the node 
corresponds to an entity.  
We insert a new tuple into \texttt{T}, assign its node and text span to 
that of $v$'s, and stop traversing $v$'s descendants.   
If no nodes evaluate to true, it implies a leaf node 
contains multiple tuples and so we flag \texttt{multi\_tuple} 
as true in \texttt{Table Catalog}
without populating \texttt{T}. 
We handle this case separately in Section~\ref{sec:query engine}.


\topic{Multi-document Extraction}
Repeated LLM calls for extracting 
tuple boundaries
for every document 
is too expensive, 
so we use a rule-based approach to populate tuples (and other
SDAs)
from the rest of the documents
that share the same template.

Consider populating \texttt{table\_node} 
for document $D' \in \mathcal{D}_i, D'\neq D$,
where tuples from $D$ were populated as described previously.  
Let the \texttt{table\_node} (i.e., the finest granularity
node below which all the tuples are found)
and \texttt{t\_range} (i.e., tuple granularity range)
of the table \texttt{T} 
in document $D$ (that has already been populated) 
be $v_{tn}$ and $[l, r]$, respectively. 
For $D'$, if there exists a node $v$ in its SHT
such that $v$'s granularity matches that
of $v_{tn}$ and the textual similarity between $v$'s phrase
and that of $v_{tn}$ is greater than a threshold,
then we set $v$ to be the table node
for $D'$; else if no such $v$ exists,
the root is set to be the table node.

Now, to populate tuples, 
suppose for the tuple range $[l, r]$ in $D, l = r = x$.
In this easy case, there is a well-defined granularity 
in the SHT where tuples are found.
Then, we add all nodes at granularity $x$ from $D'$
as candidate tuples to \texttt{T} (assuming there is a non-zero number of them).
If $l \neq r$ or if the SHT for $D'$ has a maximum height $<x$,
then we simply set \texttt{multi\_tuple} to true;
in this case, the granularity for tuples is ambiguous, 
and so we treat it similar to the case where there
may be multiple tuples at a given node.
\papertext{We describe
this step in more detail in our technical report.}

\techreport{
\topic{Multi-document Extraction Rules}
In more detail, we define the following two rules. 
For each node $v$ in an SHT, we use $v.attr$ 
to denote any attribute $attr$ belonging to $v$ in the SHT table (e.g., $v.granularity$). 
For the document $D'\in \mathcal{D}_i$, 
let $V_{D'}$ be the set of nodes corresponding to $D'$ 
in the SHT table, and $D'$\texttt{.table\_node} 
be the \texttt{table\_node} of \texttt{T} in document $D'$ in Table Catalog.}

\techreport{
\noindent\underline{Rule 1}: $\forall v_i \in V_{D'}$, if $v_i.granularity = v_{tn}.granularity$ as well as $Sim(v_i.name, v_{tn}.name > \theta$, then   $D'$\texttt{.table\_node}$ = v_i$. \\Else, $D'$\texttt{.table\_node}$ = root.$}

\techreport{
\noindent If the rule is unsatisfied, we set the \texttt{table\_node} 
to be the root node of SHT corresponding to $D'$. 
To populate the nodes corresponding to tuples, we first populate the granularity range of tuples \texttt{t\_range}.}

\techreport{
\noindent\underline{Rule 2}: If $\exists v_j \in D'$\texttt{.table\_node.child\_ids,} $l \leq v_j.granularity \leq r$, then   $D'$\texttt{.t\_range}$= [l,r]$. 
Else, \texttt{multi\_tuple} $= true$. 
}

\techreport{If the granularities of tuples of \texttt{T} 
in document $D'$ are consistent, i.e., $l=r$ in $D'$\texttt{.t\_range}, 
then we create a set of nodes $V$, 
where for each $v\in V$, $v.granularity = l$ 
and $D'$\texttt{.table\_node}$\in v.ancestor\_ids$.  
$V$ is further converted to a set of tuples whose  
$text\_span = v.context$ and $nodes = \{v\}$. 
These tuples are inserted into the  table \texttt{T}.  
If Rule 2 is violated, we set \texttt{multi\_tuple} 
as true to denote that we do not have a one-to-one 
mapping between the set of nodes and tuples when populating 
the table for $D'$. 
Note that doing so might introduce false positives instead of false negatives. 
False positives are permissive since they will not 
lose the context of where the answers may be present, 
and in Section~\ref{sec:query engine} we will discuss how to 
reduce false positives during query execution. 
When \texttt{multi\_tuple} in $D$ is true, 
we don't populate \texttt{t\_range} but set \texttt{multi\_tuple} 
as true for $D'$. 
Overall, when the number of distinct templates (i.e., $|\mathcal{D}_i|$) 
in documents $\mathcal{D}$ is small, 
the cost incurred by LLMs to populate the SDAs is minimal, 
since we only invoke LLMs on a single document for each cluster.}

\section{Query Engine}
\label{sec:query engine}

\begin{figure}[tb]
    \centering
    \papertext{\vspace{-1em}}
    \includegraphics[width=0.75\linewidth]{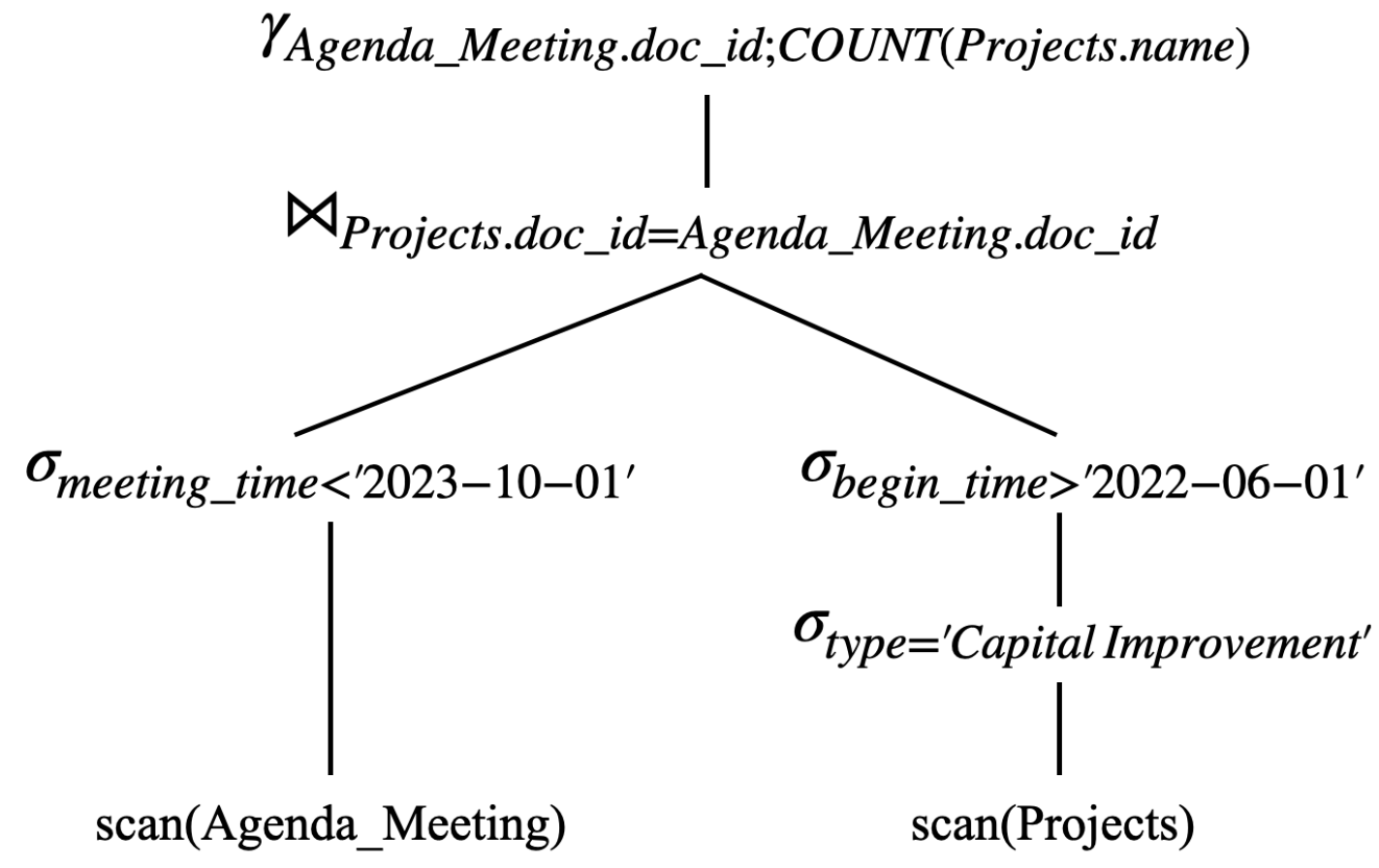}
    \vspace{-1em}
    \caption{A Query Plan for the Query in Figure~\ref{fig:query}. }
    \label{fig:queryplan}
    \papertext{\vspace{-1em}}
\end{figure}

We discuss how \sys generates a query 
plan for a given query $Q$ in Section~\ref{subsec:logicalplan}, 
and then describe 
our physical operator implementations 
that leverage SHTs in 
Section~\ref{subsec:physicalplan}.

\subsection{Logical Query Plan}
\label{subsec:logicalplan}
Unlike traditional settings
where I/O and computation
costs dominate,
here, LLM invocations 
add to monetary cost\footnote{This is common for several commercial LLMs like OpenAI, Claude-3~\cite{claude}, Google Gemini~\cite{gemini}.} and/or latency,
and thus must be minimized if possible.
Keeping this guideline in mind,
when generating a logical query plan for a 
given query $Q$,
\sys first
parses the SQL query into a parse tree
of relational operators. 
Subsequently,
predicates are pushed down
to reduce intermediate
sizes and thereby downstream LLM invocations---but
also taking into account the fact that predicate
evaluations that rely on LLMs 
can be expensive.
\sys relies on the standard approach 
from prior work~\cite{hellerstein1993predicate} for
expensive predicate reordering 
that takes into account both the
selectivity and cost. 
\papertext{Detailed descriptions can be found in our technical report~\cite{textdblong}. }
\techreport{Specifically, 
we define a metric $f(o)$ for each selection operator $o$. 
Let $s_{o}$ be the selectivity of $o$, 
computed as $s_{o} = \frac{|T_s|}{|T_c|}$, where $T_c$ ($T_s$) 
are tuples that are processed (satisfy) 
the predicate associated with $o$. 
Let $e_o$ be the average cost for 
evaluating a tuple using operator $o$, 
which is estimated adaptively 
during query execution as more tuples are processed by $o$.  
The goodness of a selection operator $o$ is then defined as 
$f_o = e_o \times s_o$. Intuitively, 
if an operator $o$ has lower cost $e_o$ 
and selectivity $c_o$, $o$ is preferred to be executed early.  \sys will sort the set of selection operators on the same table
in the increasing order of $f(o)$. }
Projections on the other hand,
are pulled up,
to avoid having to populate attributes through
LLM calls
for tuples that may get discarded.
Until a selection or projection
is encountered that requires a specific
attribute for a tuple, that attribute
stays uninterpreted, and therefore NULL.

From a join order standpoint,
\sys adopts a
greedy algorithm to generate
a left-deep tree, in an approach
akin to standard relational
query optimization techniques.
Here, instead of optimizing for reducing
the sizes of intermediate results,
we focus on reducing the LLM invocation cost.
Let $E(T)$ be the cost (in terms of dollars or latency) for
evaluating all of the predicates in $Q$ 
corresponding only to table $T$
on all of the tuples of $T$. 
\sys ranks the tables in $Q$ as $T_1, T_2, ...$  
based on their $E(T_i)$ in increasing order,
forming a left deep tree with $T_1$ as the driving table, 
followed by $T_2$ to form $T_1 \bowtie T2$, 
with the remaining tables being selected based on $E(.)$. 
When \texttt{multi\_tuple} is false, 
implying that in table $T$,
we have pre-populated potential tuples,
and therefore have a more precise estimate,
$E(T) = |T|\times e$ is estimated at query time,
where $|T|$ is the number of tuples in $T$,
$e$ denotes
the average cost of evaluating a single tuple. 
\techreport{Initially, $E(T)$ is set to be $|T|$ to 
prioritize evaluating the table with the smaller number of tuples, 
and $e$ will be estimated adaptively as more tuples are processed during query execution.}
One logical plan 
for the query in Figure~\ref{fig:query} 
is shown in Figure~\ref{fig:queryplan}, 
where \texttt{agenda\_meeting} only has one tuple 
compared to the \texttt{Projects} table 
with more than 40 tuples, and thus is evaluated first. 
The estimation of $E(T)$ when \texttt{multi\_tuple} is true 
will be described in Section~\ref{subsec:physicalplan}.

\subsection{Physical Query Plan} 
\label{subsec:physicalplan}

During query execution, 
each tuple in the user-defined DTables 
has attribute values 
that begin as \texttt{NULL} as in Figure~\ref{fig:model_query}a,
but some attributes will get populated through selections
or projections.
When \texttt{multi\_tuple} is true, \sys 
leverages LLMs to create a set of tuples 
satisfying the corresponding predicates 
with their attributes listed in the projections to be computed, 
as will discussed shortly.
We now discuss our implementations
of various operators.





\topic{Scan} 
As part of our scan operator, 
\sys executes the query document by document
(which explains the restriction of join
on \texttt{doc\_id} in Section~\ref{subsec:query language}). 
This operator first retrieves 
the tuples in the first document 
as a batch, 
followed by tuples in the second document;
thus only one SHT is processed at a time.

\setlength{\textfloatsep}{0pt}
\begin{algorithm}[bt]
    \small
	\caption{$tree\_evaluate(SHT, tuple, e)$}
	\label{alg:extractvalue}
        $CurrentNodes = \{tuple.node\}$\\
        $Ans = \emptyset$ \\ 
        $T = getTree(SHT, node)$ \\ 
        \textbf{/*Refine candidate nodes*/} \\ 
        \While{$stop\_condition(T) = False$}
        {
        $CNs = \emptyset$ \\ 
        \For{$n\in CurrentNodes$}
        {
            \If{$search\_oracle(n,e)$ = True}{  
                $CNs = CNs \cup n$ \\ }
            
        }
            $CurrentNodes = CNs.childs\_id$\\
        }
        \If{e.type = predicate}{
        \textbf{/*Evaluating A Predicate*/}\\
        \For{$node \in CNs$}{
            \If{$evaluate\_oracle(node.summary,e) = True$}{
            $Ans = Ans \cup node$ \\ 
            \textbf{Return} $Ans$
            }
        }
        }
        \If{e.type = attribute}{
            \textbf{/*Extracting Attribute Values*/}\\
            \For{$node \in CNs$}{
            $Ans = Ans \cup extract\_oracle(node.summary,e)$ \\ 
        }
        }
        \textbf{Return} $Ans$
\end{algorithm}

\topic{Selections and Projections} 
Consider a predicate $pred$ or a projection $proj$ on table $T$;
a similar procedure is followed in either case.
Say \texttt{multi\_tuple} is false,
so each row in $T$ corresponds
to a single potential tuple.
\sys then calls a function
$evaluate (SHT, tuple, e)$,
listed in Algorithm~\ref{alg:extractvalue},
with $e$ set to $pred$ (respectively, $proj$)
to evaluate whether $tuple$ satisfies $pred$,
returning it if so (respectively, the value of the attribute in $proj$). 
This function implements a tree search
on the SHTs, leveraging 
summaries for each node, as defined in Section~\ref{subsec:datamodeldef}. 
We next describe how we populate this $summary$ 
per node in the SHT table (Figure~\ref{fig:model_query}c).

\subtopic{Summary Creation}
Given the SHT for a document $D$ 
and the expression $e$,  
$S(v)$, the summary 
for a node $v$, comprises the following:
(1) The phrase(s) corresponding to both $v$ and its ancestors.  
(2) An extractive summary of the text span of $v$, 
which is a set of important sentences determined using 
standard (non-LLM) NLP tools like NLTK~\cite{nltk}.
(3) The top-1 sentence the text span of $v$ 
with the highest semantic similarity (e.g., cosine similarity) 
with $e$.

Parts (1) and (2) are prepared offline
when the SHT is built. Part (3) is added during query processing.
Including phrases (i.e., headers) of ancestors
in (1) often helps enhance accuracy by including 
additional background for interpreting $v$'s text span. 
For example, in Figure~\ref{fig:civic}, 
the summary of node $B2$ contains the 
header phrase 
of its parent, ``Capital Improvement Projects (Design)'', 
helping us identify $v$ as a candidate 
node when evaluating a predicate such as \texttt{type = Capital Improvement}.   


\subtopic{Tree Search Algorithm}
Given a document $D$ with its $SHT$, 
a tuple node $node$, an expression $e$ 
(either a predicate or a projection), 
our Algorithm~\ref{alg:extractvalue},
first identifies a sub-tree $T$ in $SHT$ with $node$ as the root (Line 4), searches $T$ top-down. 
For each node $n$ in one layer, 
it calls $search\_oracle(n,e)$ 
to check whether $n$'s summary 
contains the right information 
to evaluate expression $e$. 
It then adds all the nodes that pass 
$search\_oracle$ into a candidate set $CNs$ (Line 6-12), 
and recursively searches 
their children until a stopping condition is met (Line 6). 
This condition is (1) the leaf node is reached, 
(2) the number of tokens in the summary of the node 
is larger than that of its context (i.e., text span). 

\begin{mypython}
    search_oracle(node, e): If the following text contains the information that describes [e.descr], return True; otherwise, return False. The context is [node.summary]. 
    Example: [e.descr] = 'the type of project is Capital Improvement'
\end{mypython}

\noindent For each candidate node $n \in CNs$, 
if the expression $e$ is a predicate, 
then a call to an LLM with 
prompt $evaluate\_oracle (node.summary,e)$ 
is issued to evaluate if the summary of node satisfies the predicate. 
This step stops early when there exists one 
node that passes $evaluate\_oracle(rc,e)$ (Line 11-17). 
When $e$ is a projected attribute, $extract\_oracle(node.summary,e)$ 
is instead used to extract the value of the projected attribute (Line 18-22).  

\begin{mypython}
    evaluate_oracle(context, e):  Return True if  [e.descr] based on the following context [context]. Otherwise, return False. 
    Example: [e.descr] = 'type of project is Capital Improvement'
\end{mypython}

\begin{mypython}
    extract_oracle(context, e): Return [e.descr]  based on the following context [context]. 
    Example: [e.descr] = 'name of project'
\end{mypython}

Each selection operator $o$ returns the set of tuples in table $T$ satisfying the predicate associated with $o$ to downstream operators. 
\papertext{
We describe how \sys handles the case where \texttt{multi\_tuple} is true  for table $T$ with detailed prompts, procedure, and examples in our technical report~\cite{textdblong}.}
\techreport{
We handle the case where \texttt{multi\_tuple} is true  for table $T$ in Section~\ref{sec:multituple}.
}

Even though executing a tree search procedure
by exposing node summaries to LLMs incurs additional cost, 
it is minimal in practice since the height of the tree 
is often small\techreport{~(thus, the
number of iterations is small)}, 
and the size of the summary is small and controllable. 
In Section~\ref{sec:evaluation} we show that the benefit 
introduced by summaries, which achieves better accuracy 
and lower cost, dominates the additional cost.

\topic{Other Operators}
We use nested loop as our join algorithm.
As mentioned earlier, even if we
consider latency to be the primary optimization criterion,
the evaluation of predicates and projections
through LLM invocations
would dominate overall latency,
and the number of intermediate tuples
to be processed
during query execution is often not a large number. 
\techreport{If we further treat monetary cost as the primary criterion, 
then joins are effectively free.}
Thus, a simple nested loop join suffices.
Similarly, other operators 
like aggregation and group-by
use simple relational variants. 

\topic{Provenance of Query Answers} 
\sys maintains the provenance in the form
of the corresponding text span(s) 
for the returned query answers
in a manner analogous to classical relational provenance~\cite{glavic2021data}.
During query processing, we keep track
of the sequence of text spans consulted
to populate attributes or verify predicates,
as an additional metadata attribute, per tuple.
These text spans are combined into an array 
during joins.
While we could apply the same idea
to aggregations and capture the provenance
of contributing tuples into an array, this representation is unwieldy.
Determining how best to show all of this provenance
to end-users to ensure trust in query answers is
an important topic for future work.

\techreport{\subsection{Operators for the Multiple Tuple Case}\label{sec:multituple}}

\techreport{
When \texttt{multi\_tuple} is true for table \texttt{T}, there are no tuples in \texttt{T} after population in Section~\ref{subsec:datamodelpopulate}, 
and the context of \texttt{table\_node} may contain multiple tuples. Let $pred(T)$ and $proj(T)$ be a set of predicates and projected attributes associated with table \texttt{T} in a given query $Q$. 
In this case, \sys searches the text span corresponding to the \texttt{table\_node} of 
\texttt{T}, and creates a set of tuples satisfying $pred(T)$ with $proj(T)$ being populated by LLMs. 
}

\techreport{
When \texttt{table\_node} is a leaf node in its SHT, \sys  submits the prompt  \texttt{multi\_tuple\_oracle (table\_node}$, pred(T), proj(T)$\texttt{)} 
to LLMs to extract the projected values 
for the tuples that satisfies the given predicate $pred(T)$. } 

\ifshowblock
\begin{mypython}
    multi_tuple_oracle(node,pred(T),proj(T)): The following text describes one or more [tuple_descr]. For each [tuple_descr], if pred(T), then return [proj(T)] based on the following context [node.context]. 
    Example: 
    [tuple_descr] = 'paper'
    [predT] = 'publication year is greater than 2009 and conference is VLDB'
    [proj(T)] = 'name of paper, authors of paper' 
\end{mypython} 
\fi

\techreport{
 As an example,   
 consider a publication document $D$,
    where users want to create a table called \textit{Reference} with the schema as \{name, year\}, whose text span corresponds to the references section in a paper. Assume that in the SHT of $D$, the references section is a leaf node. In this case, \sys will not further parse the reference section into individual references, 
    but will call \texttt{multi\_tuple\_oracle()} to extract the paper name and authors per reference from VLDB whose publication year is later than 2009, directly over the references section. 
}

\techreport{
When \texttt{table\_node} 
is not a leaf node in its SHT of document $D$, let $D'$ be a document sharing the same template with $D$ and populating its 
system-defined attributes via $D$ in Section~\ref{subsec:datamodelpopulate}. Let $stop\_granularity$ be the granularity for stopping searching in Algorithm~\ref{alg:extractmultituple}, and $stop\_granularity = D.tuple\_range.l$, i.e., the smallest granularity of tuples in $D$. Note that this may introduce false positives (one node might correspond to multiple tuples) but would avoid false negatives (there will not exist 
nodes that correspond to portions of a tuple). 
\sys  executes \texttt{tree\_evaluate\_multi\_tuple} in Algorithm~\ref{alg:extractmultituple}.  
\sys starts searching the subtree of SHT with \texttt{table\_node} as the root (Line 4).   
We use the same summary-based search as in \texttt{tree\_evaluate} in Algorithm~\ref{alg:extractvalue} to refine the nodes that are related to the given query top-down layer by layer, and stop the search when the granularity of current layer exceeds $stop\_granularity$ (Line 6-12). For each node $n\in CNs$ that are related to 
the query and might contain multiple tuples, we call \texttt{multi\_tuple\_oracle} to extract the corresponding tuples (Line 13-15).  
}


\techreportsp{
\setlength{\textfloatsep}{0pt}
\begin{algorithm}[bt]
    \small
	\caption{$tree\_evaluate\_multi\_tuple$}
	\label{alg:extractmultituple}
        \KwIn{$SHT, table\_node, pred(T), proj(T), stop\_granularity$}
        $CurrentNodes = \{table\_node\}$\\
        $Tuples = \emptyset$ \\ 
        $granularity = table\_node.granularity$ \\ 
        $T = getTree(SHT, table\_node)$ \\ 
        \textbf{/*Refine candidate nodes*/} \\ 
        \While{$granularity \leq stop\_granularity$}
        {
        $CNs = \emptyset$ \\ 
        \For{$n\in CurrentNodes$}
        {
            \If{$search\_oracle(n,e)$ = True}{ 
            $CNs = CNs \cup n$ \\
         }
         }
         $granularity = granularity + 1$ \\ 
         $CurrentNodes = CNs.childs\_id$\\
        }

    $Ans = \emptyset$ \\ 
    \For{$n\in CNs$}{
        $Ans$ = $Ans$ $\cup multi\_tuple\_oracle(t,pred(T),proj(T))$
    }
        \textbf{Return} $Ans$
\end{algorithm}
}

\section{Evaluation}
\label{sec:evaluation}

In this section, we evaluate \sys over three real document
collections on accuracy, latency, and cost.

\subsection{Methodology}

\subsubsection{Data \& Query Sets}
We collected three real-world datasets (i.e., document collections):
scientific publications, civic agenda reports, and notice of violations;
details are displayed in Table~\ref{tab:datastats}. 

\begin{table}[]
\small
    \centering
    \begin{tabular}{|c|c|c|c|}
    \hline
    \textbf{Datasets} & \textbf{\# of Documents} & \textbf{Avg \# Pages}  & \textbf{Avg \# Tokens} \\ \hline
    Publication & 100 & 11.5 & 13230 \\ 
    Civic Agenda & 41 & 8.7 & 3185 \\ 
    Notice & 80 & 7.1 & 3719 \\ \hline
    \end{tabular}
    \caption{\small Characteristics of Datasets.}
    \label{tab:datastats}
    \vspace{-0.5em}
\end{table}

\begin{table*}[]
\small
\begin{adjustbox}{width=\textwidth,center}
\begin{tabular}{|c|c| c| c||c| c| c||c| c| c||c| c| c|}
\hline
\multirow{2}{*}{} & \multicolumn{3}{c||}{\textbf{Precision}} & \multicolumn{3}{c||}{\textbf{Recall}} & \multicolumn{3}{c||}{\textbf{Cost (\$) / Tokens ($\times$ 1000)}} & \multicolumn{3}{c|}{\textbf{Latency} (Seconds)} \\ \cline{2-13}
 Strategies                 & PUB   & CIVIC   & NOTICE   & PUB   & CIVIC   & NOTICE   & PUB   & CIVIC   & NOTICE    & PUB   & CIVIC   & NOTICE   \\ \hline
GPT\_single & \textbf{0.74} & 0.45 & \textbf{0.71} & 0.38 & 0.45 & \textbf{0.77} & 0.98 / 16.2 & 0.33 / 5.4 & 0.3 / 5.3 & 14.6 & 15.3 & 6.1                   \\ 
GPT\_merge & 0.63 & 0.34 & 0.66 & 0.4 & 0.45 & \textbf{0.72} & 0.8 / 13.2 & 0.2 / 3.2 & 0.2 / 3.7 & 12.9 & 7.4 & 5                   \\ 
RAG\_seq  & 0.51 & 0.12 & 0.36 & 0.38 & 0.13 & 0.38 & \textbf{0.02 / 0.4} & \textbf{0.02 / 0.29} & \textbf{0.01 / 0.18} & \textbf{3.76} & \textbf{5.1} & \textbf{1.3}                   \\ 
RAG\_tree & 0.51 & 0.2 & 0.2 & 0.38 & 0.04 & 0.17 & 0.07 / 1.2 & 0.04 / 0.66 & 0.02 / 0.35 & 10 & 8.9 & \textbf{1.3 }                 \\ 
\sys  & \textbf{0.72} & \textbf{0.73} & \textbf{0.73} & \textbf{0.53} & \textbf{0.84} & \textbf{0.74} & \textbf{0.03 / 0.56} & 0.03 / 0.53 & \textbf{0.02 / 0.25} & \textbf{4.8} & 7 & \textbf{1.7}                 \\ \hline
\end{tabular}
\end{adjustbox}
\caption{\small Average Precision, Recall, Cost / \# of Tokens and Latency of Strategies Per Query, Per Document,  in Publication (PUB), Civic Agenda (CIVIC), Notice of Violation (NOTICE) Datasets. (GPT-4-32k is Used.)}
\label{table:all}
\vspace{-1.5em}
\end{table*}

\techreport{
\begin{table*}[]
\centering
\begin{adjustbox}{width=\textwidth,center}
\begin{tabular}{|c|c c c|c||c c c|c||c c  c| c||c c c|c|}
\hline
\multirow{2}{*}{} & \multicolumn{4}{c||}{\textbf{Precision}} & \multicolumn{4}{c||}{\textbf{Recall}} & \multicolumn{4}{c||}{\textbf{Cost (\$) / Tokens ($\times$ 1000)}} & \multicolumn{4}{c|}{\textbf{Latency} (Seconds)} \\ \cline{2-17}
 Strategies                 & QG1   & QG2   & QG3   & Avg   & QG1   & QG2   & QG3   & Avg   & QG1    & QG2   & QG3   & Avg   & QG1   & QG2   & QG3   & Avg   \\ \hline
GPT\_single                 & \textbf{0.94}  & \textbf{0.66}  & \textbf{0.62}  & \textbf{0.74}   & 0.65  & 0.16  & 0.32  & 0.38   & 0.8 / 13.2   & 1 / 16.6  & 1.1 / 18.9  & 0.98 / 16.2   & 12.8  & 14.1  & 16.9  & 14.6   \\ 
GPT\_merge                 & \textbf{0.94}  & 0.41  & 0.63  & 0.63   & 0.65  & 0.13  & 0.41  & 0.4   & 0.8 / 13.2   & 0.8 / 13.2  & 0.8 / 13.2  & 0.8 / 13.2   & 12.8  & 12.9  & 13.1  & 12.9   \\ 
RAG\_seq                 & 0.73  & 0.4  & 0.39  & 0.51   & 0.6  & 0.23  & 0.31  & 0.38   & \textbf{0.01 / 0.23 }  & \textbf{0.02 / 0.38}  & \textbf{0.03 / 0.59}  & \textbf{0.02 / 0.4}   & \textbf{2.7} & \textbf{3.9} &  \textbf{4.5} & \textbf{3.76}   \\ 
RAG\_tree                 & 0.79  & 0.33  & 0.42  & 0.51   & \textbf{0.68}  & 0.19  & 0.27  & 0.38   & 0.05 / 0.82   & 0.08 / 1.3  & 0.1 / 1.6  & 0.07 / 1.2   & 8.4 & 10.4 & 11.2 & 10  \\ 
\sys                 & \textbf{0.93}  & \textbf{0.64}  & \textbf{0.6}  & \textbf{0.72}   & \textbf{0.7}  & \textbf{0.54}  & \textbf{0.34}  & \textbf{0.53}   & \textbf{0.02 / 0.41}   & \textbf{0.03 / 0.56}  & \textbf{0.04 / 0.71}  & \textbf{0.03 / 0.56}   & \textbf{3.9} & \textbf{5.2} &  \textbf{5.4} & \textbf{4.8}  \\ \hline
\end{tabular}
\end{adjustbox}
\caption{\small Average Precision, Recall, Cost / \# of Tokens and Latency of Strategies Per Query, Per Document,  in Publication Dataset. (GPT-4-32k is Used.)}
\label{table:publication}
\vspace{-1em}
\end{table*}
}

\techreport{
\begin{table*}[]
\centering
\begin{adjustbox}{width=\textwidth,center}
\begin{tabular}{|c|c c c|c||c c c|c||c c  c| c||c c c|c|}
\hline
\multirow{2}{*}{} & \multicolumn{4}{c||}{\textbf{Precision}} & \multicolumn{4}{c||}{\textbf{Recall}} & \multicolumn{4}{c||}{\textbf{Cost (\$) / Tokens ($\times$ 1000)}} & \multicolumn{4}{c|}{\textbf{Latency} (Seconds)} \\ \cline{2-17}
 Strategies                 & QG1   & QG2   & QG3   & Avg   & QG1   & QG2   & QG3   & Avg   & QG1    & QG2   & QG3   & Avg   & QG1   & QG2   & QG3   & Avg   \\ \hline
GPT\_single                 & 0.64  & 0.36  & 0.36  & 0.45   & 0.73  & 0.37  & 0.24  & 0.45   & 0.2 / 3.2   &  0.33 / 5.4  & 0.47 / 7.6  & 0.33 / 5.4   & 7.5 & 15.3 &  23.1 & 15.3   \\ 
GPT\_merge                 & 0.64  & 0.22  & 0.16  & 0.34   & 0.73  & 0.32  & 0.29  & 0.45   & 0.2 / 3.2   & 0.2 / 3.2  & 0.2 / 3.2  & 0.2 / 3.2   & 7.3  & 6.9  & 7.5  & 7.4   \\ 
RAG\_seq                 & 0.25  & 0.11  & 0  & 0.12   & 0.36  & 0.04  & 0  & 0.13   & \textbf{0.01 / 0.14}   & \textbf{0.02 / 0.3}  & \textbf{0.03 / 0.43}  & \textbf{0.02 / 0.29}   & \textbf{3.3} & \textbf{5.2} &  \textbf{6.9} & \textbf{5.1}   \\ 
RAG\_tree                 & 0.36  & 0.23  & 0  & 0.2   & 0.12  & 0.01  & 0  & 0.04   & 0.03 / 0.49   & 0.04 / 0.6  & 0.05 / 0.88  & 0.04 / 0.66   & 5.9 & 8.9 & 12.3 & 8.9  \\ 
\sys                 & \textbf{0.89}  & \textbf{0.72}  & \textbf{0.61}  & \textbf{0.73}   & \textbf{0.86}  & \textbf{0.79}  & \textbf{0.83}  & \textbf{0.84}   & 0.02 / 0.43   & 0.04 / 0.59  & 0.04 /  0.68  & 0.03 / 0.53   & 5.1 & 7.2 &  8.8 & 7  \\ \hline
\end{tabular}
\end{adjustbox}
\caption{\small Average Precision, Recall, Cost / \# of Tokens and Latency of Strategies Per Query, Per Document,  in Civic Dataset. (GPT-4-32k is Used.)}
\label{table:civic}
\vspace{-1em}
\end{table*}
}

\techreport{
\begin{table*}[]
\centering
\begin{adjustbox}{width=\textwidth,center}
\begin{tabular}{|c|c c c|c||c c c|c||c c  c| c||c c c|c|}
\hline
\multirow{2}{*}{} & \multicolumn{4}{c||}{\textbf{Precision}} & \multicolumn{4}{c||}{\textbf{Recall}} & \multicolumn{4}{c||}{\textbf{Cost (\$) / Tokens ($\times$ 1000)}} & \multicolumn{4}{c|}{\textbf{Latency} (Seconds)} \\ \cline{2-17}
 Strategies                 & QG1   & QG2   & QG3   & Avg   & QG1   & QG2   & QG3   & Avg   & QG1    & QG2   & QG3   & Avg   & QG1   & QG2   & QG3   & Avg   \\ \hline
GPT\_single                 & \textbf{0.71}  & \textbf{0.65}  & \textbf{0.76}  & \textbf{0.71}   & \textbf{0.9}  & \textbf{0.67 } & \textbf{0.75}  & \textbf{0.77}   & 0.2 / 3.7   & 0.31 / 5.2  & 0.43 / 7.1  & 0.3 / 5.3   & 4.9 & 6.2 &  7.3 & 6.1   \\ 
GPT\_merge                 & \textbf{0.7}  & 0.56  & 0.62  & 0.66   & 0.8   & 0.6  & \textbf{0.77}  & \textbf{0.72}   & 0.2 / 3.7   & 0.2 / 3.7  & 0.2 / 3.7  & 0.2 / 3.7   & 4.8  & 5  & 5.1  & 5   \\ 
RAG\_seq                 & 0.61  & 0.31  & 0.17  & 0.36   & 0.67  & 0.22  & 0.26  & 0.38   & \textbf{0.01 / 0.12}   & \textbf{0.01 / 0.19}  & \textbf{0.01 / 0.23}  & \textbf{0.01 / 0.18}   & \textbf{0.9} & \textbf{1.3} &  \textbf{1.7} & \textbf{1.3}   \\ 
RAG\_tree                 & 0.58  & 0.36  & 0.24  & 0.2   & 0.39  & 0.5  & 0.17  & 0.17   & 0.02 / 0.25   & 0.02 / 0.38  & 0.03 / 0.41  & 0.02 / 0.35   & 2.1 & 2.7 & 3.1 & 2.6  \\ 
\sys                 & \textbf{0.79}  & \textbf{0.67}  & \textbf{0.72}  & \textbf{0.73}   & \textbf{0.87}  & 0.62  & \textbf{0.73}  & \textbf{0.74}   & \textbf{0.01 / 0.19}   & \textbf{0.02 / 0.26}  & \textbf{0.02 / 0.3}  & \textbf{0.02 / 0.25}   & 1.4 & 1.7 &  2.1 & 1.7  \\ \hline
\end{tabular}
\end{adjustbox}
\caption{\small Average Precision, Recall, Cost / \# of Tokens and Latency of Strategies Per Query, Per Document,  in Notice Violation Dataset. (GPT-4-32k is Used.)}
\label{table:notice}
\vspace{-1em}
\end{table*}
}

\noindent\textbf{Scientific Publications.} 
This dataset was collected from a systematic review study 
that examined research questions in 
the field of personal data management at UC Irvine~\cite{paperdata}. 
The study analyzed over 500 publications; 
we randomly selected 100 papers for our dataset. 
The study explored 20 research questions 
with human-labeled answers for all of the publications. 


\noindent\textbf{Civic Agenda Reports.} 
This dataset, from our collaborators at Big Local News, 
comprises 41 civic agenda reports from 2022 to 2024 
in the City of Malibu~\cite{civicdata}. 
Each report details a series of government projects, 
including their status, updates, decisions, 
and timelines for beginning, ending, and expected construction. 


\noindent\textbf{Notice of Violations.} 
This dataset, also from Big Local News, of 80 documents
describe notices of violations issued by the US Dept.~of Transportation 
from 2023 to 2024~\cite{noticedata}. 
Each document concerns potential violations detailed by the Hazardous Materials Safety Administration,
including detailed violation orders and descriptions, penalty decisions, 
and proposed compliance orders. 


\noindent\textbf{Query Workload.} 
For each dataset, we 
devise a query workload comprising 9 SQL queries,
informed by the needs of our collaborators. 
These 9 queries are divided into groups of three,
QG1, QG2, and QG3, varying in the number of predicates,
from one to three respectively.
\papertext{We present the fine-grained breakdown
by QGi in our technical report~\cite{textdblong}
and only present average results here.}
To generate these queries, we first
define tables along with a set
of attributes per dataset.
Then we randomly select
$i$ attributes to create $i$ predicates
for the queries in group QGi,
and in \texttt{SELECT}, we 
additionally include one 
attribute that is not used in the predicates,
as well as \texttt{doc\_id}.
When we end up sampling attributes
across multiple relations, we list
both in the \texttt{FROM} clause,
and additionally add an equijoin condition
on \texttt{doc\_id}.
So, overall, our queries include selections, projections, and joins.
We omit aggregations in our workload since
we use relational versions for those operators
evaluated after the corresponding attribute values are extracted;
and thus the performance on such queries would be similar
to that on the queries without them.

\subsubsection{Strategies Compared and Evaluation Metrics. } 
We compare \sys with four baselines, 
GPT\_single, GPT\_merge, RAG\_seq, and RAG\_tree.  
The first two operate
on an entire document at a time.
GPT\_single uses a separate LLM call per predicate
and projection by constructing a corresponding prompt,
appending the entire document as context. 
GPT\_merge combines all of the predicates and projections
into a single LLM call alongside the entire document.
RAG\_seq and RAG\_tree refer to RAG-based techniques
in two variants implemented by LlamaIndex~\cite{llamaindex},
a state-of-the-art open-source RAG framework: 
sequential chunking and tree-style chunking, respectively.  
In RAG\_seq, 
we set the chunk size to 128 tokens 
and selected top-$k$ chunks, 
where $k = \max(1,5\% \times \texttt{doc\_size}/128)$.
That is, we retrieve at least one chunk, but no more than 5\% of the
of the document. 
RAG\_tree constructs a hierarchical tree from the document
without leveraging semantic structure. 
This tree is constructed by first chunking 
the leaves at a fixed granularity.
Nodes higher up in the hierarchy
are formed by recursively summarizing 
the nodes below.
Subsequently, a path from the root to leaf
is retrieved, instead of just one leaf. 
GPT-4-32k is used to evaluate the queries for all strategies. 
\papertext{Additional details of prompts for the 
baselines can be found in our technical report~\cite{textdblong}.}

We use precision and recall to measure the quality of query answers. 
Given a query $Q$,  let $T\_truth(Q)$ and $T\_pre(Q)$ 
be the set of tuples in the ground truth vs.~predicted by
an approach, respectively. 
Precision is measured as $\frac{|T\_truth(Q)\cap T\_pre(Q)|}{|T\_pre(Q)|}$, 
and  recall is $\frac{|T\_truth(Q)\cap T\_pre(Q)|}{|T\_truth(Q)|}$. 
We count the number of input and output tokens
to measure the cost of LLM invocations~\cite{gpt-4}. 
Finally, we measure the latency of query execution 
by taking three runs and reporting the average.  

\begin{figure*}
\centering
\subfigure[Average Precision]{
\label{fig:summary_precision}
\includegraphics[width=4.2cm,height=3cm]{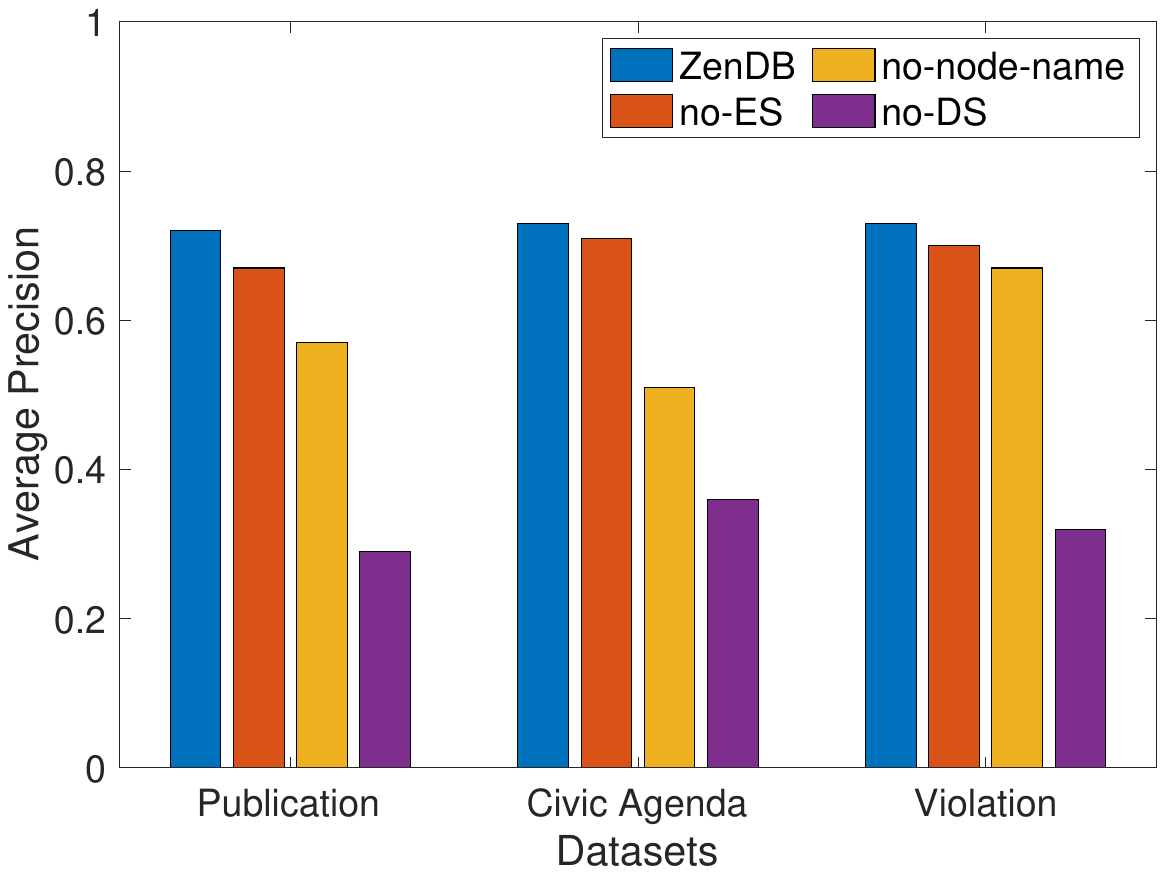}}
\subfigure[Average Recall]{
\label{fig:summary_recall}
\includegraphics[width=4.2cm,height=3cm]{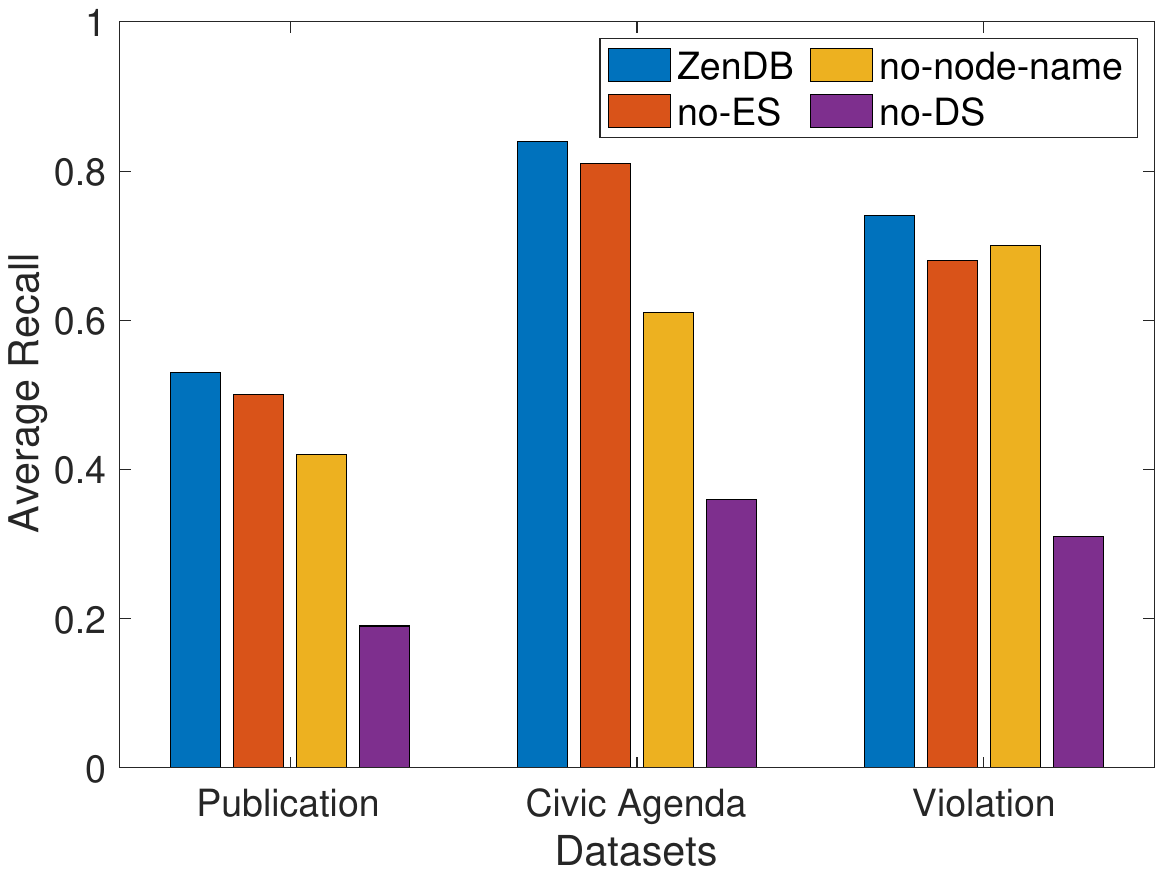}}
\subfigure[Average \# of Tokens ($\times$ 1000)]{
\label{fig:summary_cost}
\includegraphics[width=4.2cm,height=3cm]{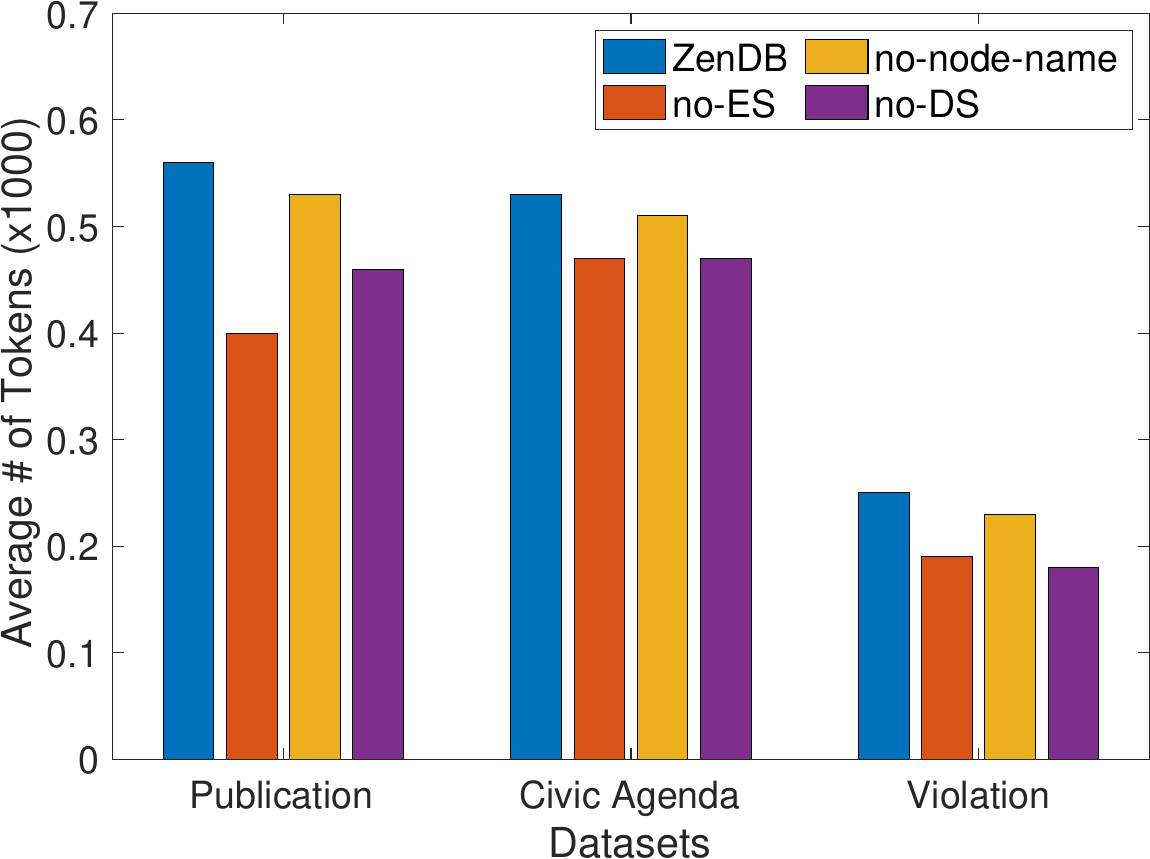}}
\subfigure[Average Latency (Seconds)]{
\label{fig:summary_latency}
\includegraphics[width=4.2cm,height=3cm]{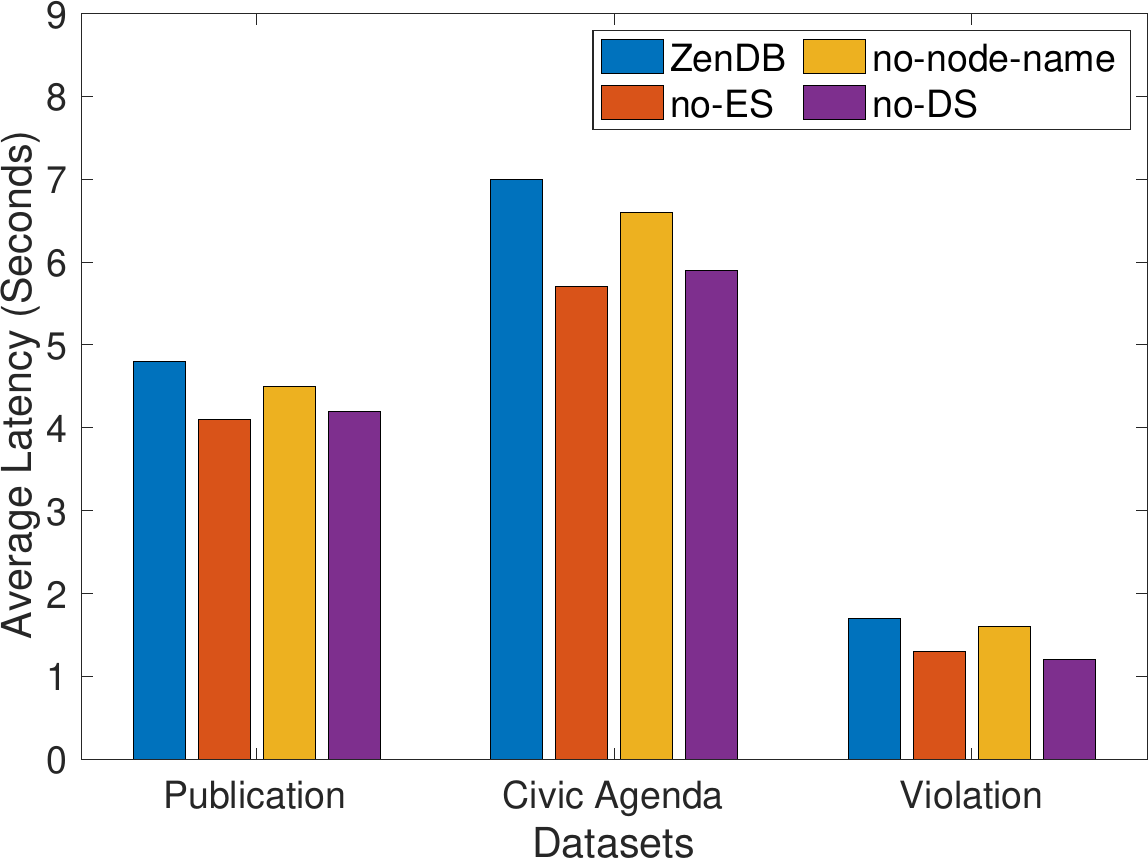}}
\vspace{-1.5em}
\caption{\small The Effect of Summary Construction to Performance of \sys in Real Datasets. }
\vspace{-1em}
\label{fig:summary}
\end{figure*}
\subsection{Experimental Results}
\topic{Experiment 1: \sys vs. GPT-only  Strategies} 
\papertext{
We first compare \sys with GPT\_single and GPT\_merge,
both operating on an entire document at a time. 
Table~\ref{table:all} reports our metrics of interest
on the three datasets.}
\techreport{
We first compare \sys with GPT\_single and GPT\_merge,
both operating on an entire document at a time. 
Table~\ref{table:all} reports our metrics of interest
on the three datasets, while Table~\ref{table:publication}, Table~\ref{table:civic}, and  Table~\ref{table:notice} provide a breakdown per dataset.}
We first note that 
\sys achieves comparable precision 
and recall to GPT\_single on the publication and notice datasets. 
Notably, \sys surpasses GPT\_single in the civic dataset, 
{\bf \em improving precision by 28\% and recall by 39\%},
due to this dataset's complex semantic structure, 
which poses challenges for GPT\_single 
in generating high-quality responses. 
\sys's approach of querying based on SHTs,
focuses LLM attention on portions of documents at a time, thereby enhancing performance. 
We also observe that combining multiple predicates 
into a single prompt 
makes it more difficult for the LLM to provide
the correct answer, 
resulting in performance degradation. 
On the cost and latency front, \sys significantly reduces both relative
to GPT\_single and GPT\_merge. 
Specifically, \sys achieves {\bf \em cost savings of approximately  29$\times$, 10$\times$, and 4$\times$} 
for the publication, civic, and notice datasets respectively. 
It's noteworthy that \sys's cost savings 
increase with document size, 
as the number of tokens it uses 
is somewhat independent of document size. 
Instead, it relies on the size of the summary 
and the number of levels of the SHTs 
explored during execution, 
which are controllable factors. 
Accordingly, we observe varying levels of latency savings with \sys, 
{\bf \em up to a 4$\times$ reduction} across datasets. 

\topic{Experiment 2: \sys vs. RAG-only Strategies} 
When compared with RAG\_seq and RAG\_tree, 
we observe that RAG\_seq achieves 
significant cost and latency savings compared to GPT-only strategies. 
However, relying solely on retrieving
physical chunks based on 
embedding similarity as in RAG,
fails to accurately identify 
the appropriate text spans related to the queries, 
leading to a substantial degradation in precision and recall. 
While \sys incurs a slightly higher cost, 
it offers substantial advantages over RAG-based approaches 
thanks to the use of semantic structure,
with {\bf \em increases in precision by up to 61\% and recall by up to 80\%}. 
RAG\_tree generally shows slight improvements in precision and recall over RAG\_seq, 
but it similarly falls short of \sys for a similar reason. 
Its use of tree-style physical chunking 
often fails to accurately identify 
the appropriate text spans. 
Moreover, the exhaustive summary construction 
and usage in RAG\_tree 
results in higher cost and latency compared to \sys.



\begin{table}[]
\small
    \centering
    \scalebox{0.85}{
    \begin{tabular}{|c|c|c|c|c|}
    \hline
    \textbf{Datasets (\# of Docs)} & \textbf{\# of Nodes} & \textbf{\# of Layers} & \textbf{Cost (\$) /Tokens} & \textbf{Latency}  \\ \hline 
    Publication (100) & 13.4 & 2.8 & 0.05 / 1.8k & 6min  \\ \hline
    Civic (41) & 32.1 & 2.9 & 0.01 / 0.36k & 1min  \\ \hline
    Violation (80) & 8.9 & 2.2 & 0.01 / 0.32k & 1min  \\
    \hline
    \end{tabular}}
    \caption{\small SHT Construction. (GPT-4 is Used.)}
    \vspace{-2em}
    \label{tab:sht}
\end{table}


\topic{Experiment 3: Data Preparation} 
Next, we examine two phases within \sys
happening prior to queries, SHT construction and table population,
and compare it to the costs of online queries. 

\subtopic{Experiment 3.1: SHT Construction} 
We present the average number of nodes and layers per SHT, 
and the \textit{total} cost, number of tokens, 
and latency on three datasets, in Table~\ref{tab:sht}. 
SHT construction is an offline process, making latency at the level of minutes not problematic. 
The cost is affected by the number of distinct templates in the datasets. 
\sys uses LLMs to verify headers for SHT generation for one document per template, 
with the remaining SHTs created through visual pattern matching. 
The cost is further reduced by sampling the phrase clusters.  
In the publication dataset, the publications originate from 6 conferences, 
whereas the other two datasets follow a consistent template. 
Therefore, the publication dataset 
has a higher cost than the others, although all costs are minimal.


\begin{table}[]
\small
    \centering
    \scalebox{0.85}{
    \begin{tabular}{|c|c|c|c|c|}
    \hline
    
    \textbf{Datasets} & \textbf{Cost (\$) / Tokens} & \textbf{FP} & \textbf{FN} & \textbf{Latency}  \\ \hline
    Publication (100) & 0.048 / 95.4k & 0 & 0 & 7 min  \\ \hline
    Civic (41) & 0.005 / 10.1k & 0.08 & 0 & 3 min  \\ \hline
    Violation (80) & 0.005 / 8.9k & 0.04 & 0 & 2 min  \\
    \hline
    \end{tabular}}
    \caption{\small Table Population. (GPT-3.5-Turbo is Used.)}
    \vspace{-1em}
    \label{tab:tablepop}
\end{table}

\begin{figure}
\subfigure[\small Precision]{\includegraphics[width=4.2cm,height=3cm]{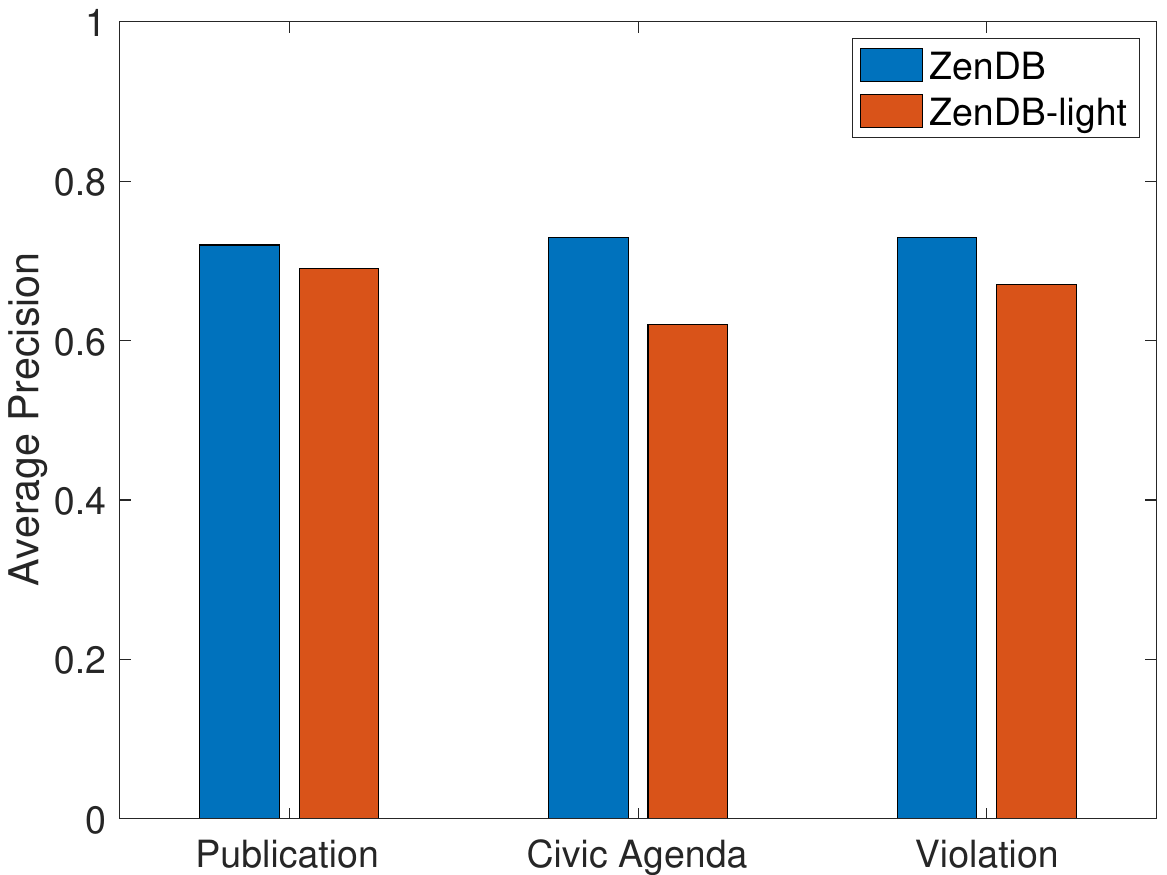}}
	\subfigure[\small Recall]{\includegraphics[width=4.2cm,height=3cm]{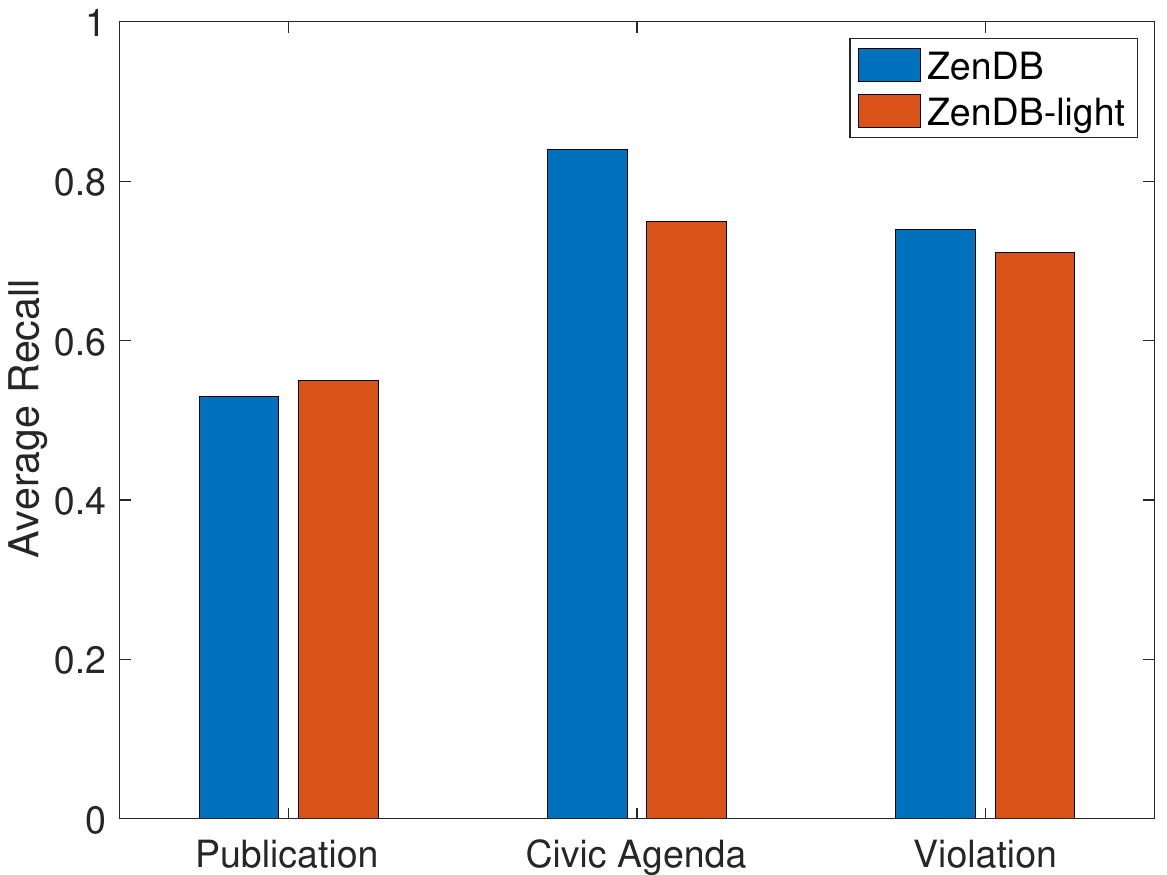}}
	\vspace{-2em}
\caption{\small \sys vs. \sys-light: Precision and Recall. } 
\label{fig:lightdb_quality}
\end{figure}

\begin{figure}
	\centering
	\begin{minipage}[t]
{0.49\linewidth}
\vspace{-0.5em}\subfigure{\includegraphics[width=4.2cm,height=3.2cm]{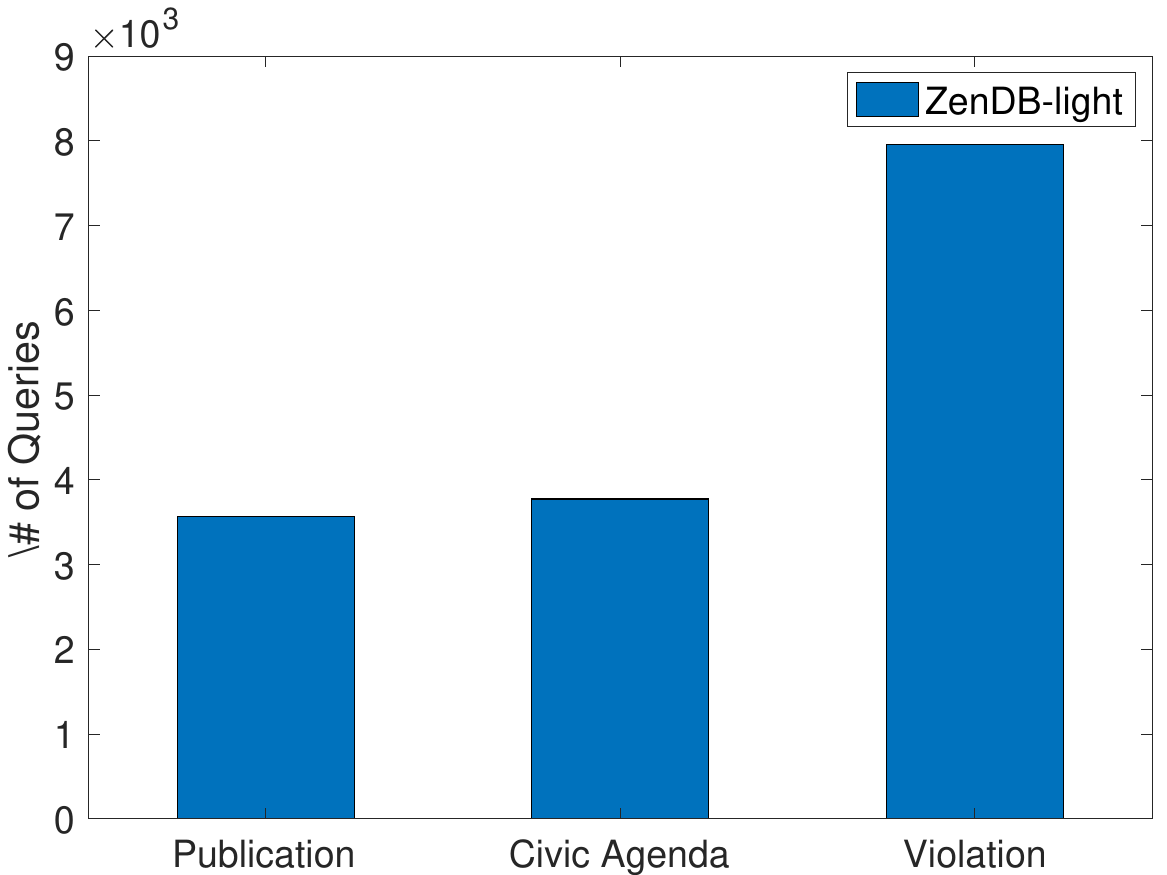}}
 \vspace{-2.1em}
		\caption{\small \# of Queries on 1 Document by 1 \$.} 
		\label{fig:lightdb_cost}
	\end{minipage}
 \hfill
 \begin{minipage}[t]{0.49\linewidth}\subfigure{\includegraphics[width=4.2cm,height=3cm]{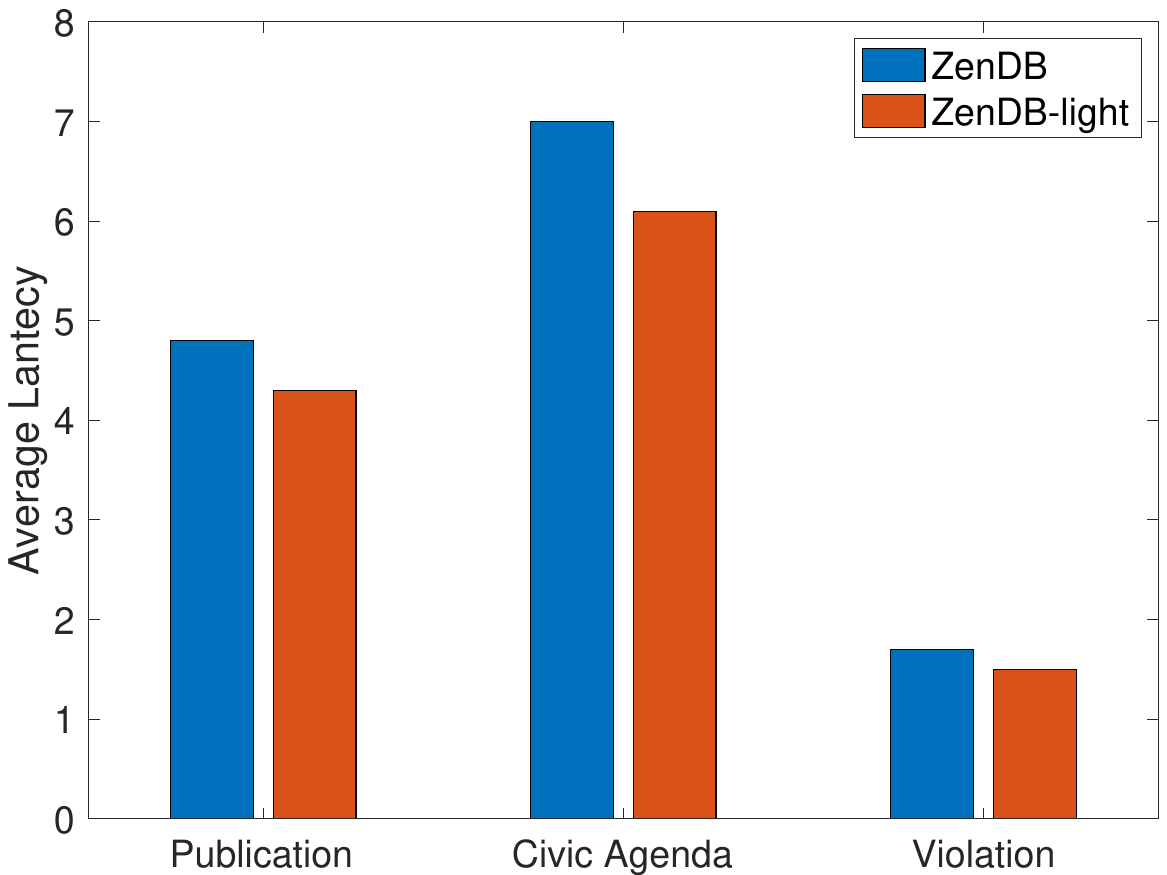}}
 \vspace{-2em}
		\caption{\small \sys VS \sys-light: Latency.}
		\label{fig:lightdb_latency}
	\end{minipage}
\end{figure}

\subtopic{Experiment 3.2: Table Population} 
When users define a DTable, 
\sys populates the system-defined attributes 
using LLM-based and rule-based approaches. 
Table~\ref{tab:tablepop} presents the \textit{total} 
cost and number of tokens (we use GPT-3.5-Turbo)\footnote{When the context size of a node exceeds the token limit (e.g., the root node in  publication dataset), we use NLTK~\cite{nltk} to summarize the context and adjust the summary size to approximately match the token limit of a prompt.}, 
with additional latency and quality results. 
In particular, to show the quality of table population, 
let $ts_g(e)$ and $ts_p(e)$ be the text span of an entity $e$ (a table or a tuple)
in the ground truth and predicted by \sys, respectively. 
We label $ts_g(e) \subset ts_p(e)$ as a false positive (FP), 
indicating that the predicted text span contains the true text span but is larger, 
which is acceptable since it doesn't miss the correct answers 
will be refined by the tree-search algorithm. 
In constrast,  $(ts_p(e) \subset ts_g(e)) \lor (ts_p(e) \cap ts_g(e)) = \emptyset)$ 
is considered a false negative (FN) because the predicted text span 
does not encompass all the true text spans, potentially resulting in missed answers. 
Notably, \sys demonstrates a low FP rate in the violation and civic agenda datasets, showcasing the effectiveness of the approach. 
The cost incurred in this step is minimal, thanks to the use of the affordable LLM GPT-3.5-Turbo (around 100x cheaper than GPT-4). 

\subtopic{End-to-end cost comparison: \sys vs. Others} 
Although \sys incurs costs to construct SHTs and 
populate tables before a query arrives, 
these costs are minimal, totaling 0.1, 0.015, and 0.015 dollars for the publication, 
civic agenda, and notice of violations datasets, 
respectively. 
{\bf \em Even if we ran just a single query subsequently, we would have lower end-to-end costs}
for \sys compared to GPT-single and GPT-merge, 
with the loading costs getting amortized across queries. 

\topic{Experiment 4: The Effect of Summary Construction in \sys}
We examine the effect of summary construction on \sys performance in Figure~\ref{fig:summary}. 
Recall that in Section~\ref{subsec:physicalplan}, the summary of each node $v$ in a SHT 
consists of three components: an extractive summary (ES), 
the phrases of $v$ and its ancestors (node-name), and the top-1 sentence 
related to a given query predicate or projection within the text span of $v$ 
(DS, i.e., Dynamic Summary). 
We explored three variations of \sys 
by removing one component at a time from the summary: 
no-ES, no-node-name, and no-DS\techreport{~(e.g., no-ES refers to the strategy that excludes the extractive summary from the summary of the node)}. 
We observe that the extractive summary 
impacts the quality of query answers (i.e., precision and recall) 
the least, while both dynamic summaries 
and node names (i.e., the header phrases) 
affect performance more significantly. 
Node names provide useful metadata 
that adds more context for the LLM, helping refine the search space. 
The dynamic summary plays a critical role 
in summary construction by not only identifying the relevant nodes 
but also retrieving the text span most related to the given query. 
We also note that storing node names 
has a minimal impact on cost and latency due to their compact size. 
In contrast, both extractive and dynamic summaries have a greater size, 
though they still represent a relatively small portion 
of the overall cost and latency.



\topic{Experiment 5: \sys Driven by A Cheaper LLM: GPT-3.5-Turbo} 
We next study the impact of replacing the more expensive LLM used in \sys, GPT-4-32k, 
with an almost 100$\times$ cheaper LLM, GPT-3.5-turbo, when evaluating queries. 
We denote this version as \sys-light. 
In Figure~\ref{fig:lightdb_quality}, {\bf \em \sys-light exhibits 
approximately a 7\% decrease in precision and a 3\% decrease 
in recall compared to \sys, at 100$\times$ lower cost}. 
This demonstrates that by refining the text span 
that \sys uses for evaluating queries, as opposed to the 
entire complex document, \sys is able to provide
a much simpler and more precise context for LLMs to evaluate. 
This makes it easier for less-advanced but cheaper models like GPT-3.5-turbo to not
just process the entire text span, but also answer the query accurately. 
We report the average number of SQL queries 
that can be executed on a single document 
by spending 1 dollar using \sys-light, in Figure~\ref{fig:lightdb_cost}. 
\sys-light can run approximately 3.5k, 3.7k, and 8k SQL queries 
with 2 predicates and one projection on average in one document within 
budget for the publication, civic agenda reports, and notices of violations, respectively, 
demonstrating the practicality of \sys-light. 


\section{Related Work}
\label{sec:related}

We now survey related work on querying unstructured data.

\topic{Text-to-Table Extraction}
One approach to querying unstructured data is
by simply extracting 
unstructured data into tables,
following which they are queried as usual.
This approach is followed by Google DocumentAI~\cite{documentai} 
and Azure Document Intelligence~\cite{azuredocument},
as well as approaches such as text-to-table~\cite{wu2021text}.
Using an LLM to populate entire
tables upfront can be expensive and error-prone
on large and complex document collections as in our case.
Evaporate~\cite{arora2023language} uses an LLM to infer schema,
and then populate tables, using synthesized rules if possible.
Simple extraction rules, 
such as ones generated by Evaporate, are not applicable
in our setting.

\topic{Retrieval-Augmented Generation (RAG)}
 RAG techniques~\cite{li2022survey,izacard2022few,cai2022recent,wang2023learning}, 
help identify smaller text portions 
that are most relevant to a given query in order to fit into finite context windows, reduce cost, and in some cases improve accuracy.
Most techniques use fixed granularity chunking policies and don't account for semantic structure,  
while recent extensions rely on potentially expensive recursive summarization to build a hierarchy~\cite{sarthi2024raptor,llamaindex}.
We showed that this RAG\_tree 
approach suffers from the same issues
as vanilla RAG.  The leaf nodes still
use fixed size chunks that are divorced from semantics, and thus fail to find relevant text segments.  
In comparison, \sys leverages semantic structure
to boost precision and recall by up to 61\% and 80\%.

\topic{Multi-Modal Databases}
Recent work creates of
multi-modal databases~\cite{jo2023demonstration,chen2023symphony,urban2023towards,thorne2021natural,thorne2021database} that support SQL-like interfaces over text, images, and/or video.
However, they all apply  LLMs or other pre-trained models
to entire documents at a time, and are thus limited to simple, small documents. This is equivalent
to our vanilla LLM approach, which is expensive and not very accurate. Other work~\cite{hattasch2023wannadb} has used interactive query processing to improve query results through user feedback. 
None of these approaches have explored 
the use of semantic structure to 
reduce cost and improve accuracy. 

\topic{Natural Language Interfaces to Data}
Supporting natural language querying over structured
data is a long-standing question in the database community;
a recent survey is one by Quamar et al.~\cite{quamar2022natural}.
While the database community has been working on this problem
for over a decade, e.g., \cite{li2014nalir},
LLMs have dominated recent benchmarks~\cite{chang2023dr,li2024can}.
In our work, we instead focus on the inverse problem of structured (SQL)
queries over unstructured data---but this line of work could
aid the first step of SQL query construction.

\topic{LLMs meet Data Management}
LLMs potentially disrupts the field of data management~\cite{fernandez2023large}, but the first step is to actually understand tables.
Recent work~\cite{fang2024large,cong2023observatory,yin2020tabert} 
explores how well LLMs understand tabular data, and representing knowledge learned by the LLM as structured data~\cite{saeed2023querying,urban2023omniscientdb}.
Many data management problems have been revisited,
including query rewriting~\cite{liu2024query}, database tuning~\cite{trummer2022db}, 
data preprocessing~\cite{zhang2023large},
data and join discovery~\cite{dong2022deepjoin,kayali2023chorus,deng2022turl}, data profiling~\cite{huang2024cocoon},
and
data wrangling~\cite{narayan2022can,chen2023seed,li2020deep}.
Some recent work has also explored how well LLMs can generate tables~\cite{tang2023struc}.
\sys also uses LLMs, but to a new setting: document analytics.

\topic{Structured Extraction}
Structured extraction from web pages, pdfs, and images has a long history of work.  For instance, Snowball~\cite{agichtein2000snowball} proposed structured extraction over the open web, and leverage common techniques such as wrapper induction~\cite{kushmerick2000wrapper,manabe2015extracting} which also leverage the hierarchical structure of HTML documents and headings.  In contrast, \sys takes as input PDFs, which are often not hierarchically encoded.  Other works, such as Shreddr~\cite{chen2012shreddr} extract from images of forms where the templates are identical, and focus on efficient use of crowd workers.  These are also relevant due to the similarities between LLMs and crowdsourcing~\cite{parameswaran2023revisiting}.





\section{Conclusion}
\label{sec:conclusion}


We presented \sys, a document analytics system
that leverages templatized structure present
in documents in a collection to support
cost-efficient and accurate query processing.
During ingest, \sys extracts structure
from documents in the form of SHTs,
guaranteeing that the results are correct
for well-formatted documents.
Then, during table creation, \sys
maps tuples to nodes in the SHT, with attribute values
to be populated during querying.
\sys supports SQL queries on user-defined
document tables, applying predicate reordering and pushdown,
and projection pull-up techniques,
coupled with a summary-based tree-search approach
to optimize query processing.
Across multiple domains, \sys provides
a compelling trade-off point relative
to LLM-only or RAG based approaches. 
In future work, we plan to study the setting
where there are no templates or when the templates
are very noisy,
as well as expand the space of SQL queries
supported. In addition, we envision a rich design space
for user interfaces to allow users to explore the results
of \sys queries alongside their provenance. 


\iftoggle{fullversion}{
\input{appendix}
}
{}
\bibliographystyle{ACM-Reference-Format}
\bibliography{refs}

\end{document}